\documentclass[%
reprint,
superscriptaddress,
showkeys,
nofootinbib,
amsmath,amssymb,
aps,
pra,
]{revtex4-2}

\usepackage{graphicx}
\usepackage{dcolumn}
\usepackage{bm}
\usepackage{hyperref}
\usepackage{verbatim}
\usepackage[utf8]{inputenc}
\usepackage{layouts}
\usepackage{multirow}
\usepackage[dvipsnames]{xcolor}
\usepackage{hhline} 
\usepackage[separate-uncertainty=true]{siunitx}
\usepackage[version=3]{mhchem} 
\usepackage{wrapfig}
\usepackage{braket}
\usepackage{float}
\usepackage{leftidx} 
\usepackage{amsmath,mathtools}
\usepackage{amssymb}
\usepackage{chngcntr}
\renewcommand{\figurename}{\textbf{Figure}}

\newcommand{\moire}{moir\'e }

\newcommand{\xin}{2s^{\text{-}}_{\text{I}}}
\newcommand{\xip}{2s^{\text{+}}_{\text{I}}}

\hypersetup{colorlinks=true,urlcolor=magenta,linkcolor=magenta,citecolor=cyan}
\allowdisplaybreaks

\newcommand{\WSI}{\affiliation{Walter Schottky Institute and TUM School of Natural Sciences, Technical University of Munich, 85748 Garching, Germany}}

\newcommand{\MCQST}{\affiliation{Munich Center for Quantum Science and Technology (MCQST), 80799 Munich, Germany}}

\begin{document}

\title{Engineering strong correlations in a perfectly aligned dual \moire system}

\author{Amine Ben Mhenni}
\email{Amine.Ben-Mhenni@tum.de}
\WSI
\MCQST

\author{Elif Çetiner}%
\WSI
\MCQST

\author{Kenji Watanabe}
\affiliation{
 Research Center for Electronic and Optical Materials, National Institute for Materials Science, 1-1 Namiki, Tsukuba 305-0044, Japan
}

\author{Takashi Taniguchi}
\affiliation{
 Research Center for Materials Nanoarchitectonics, National Institute for Materials Science,  1-1 Namiki, Tsukuba 305-0044, Japan
}

\author{Jonathan J. Finley}%
\email{jj.finley@tum.de}
\WSI
\MCQST

\author{Nathan P. Wilson}%
\email{Nathan.Wilson@tum.de}
\WSI
\MCQST

\date{\today}

\begin{abstract}

Exotic collective phenomena emerge when bosons strongly interact within a lattice \cite{baranov_condensed_2012, bloch_quantum_2012, landig_quantum_2016, browaeys_many-body_2020}. However, creating a robust and tunable solid-state platform to explore such phenomena has been elusive \cite{zhang_correlated_2022, gu_dipolar_2022, lagoin_mott_2022, lagoin_extended_2022, xiong_correlated_2023, ben_mhenni_gate-tunable_2024, gao_excitonic_2024}.
Dual \moire systems---compromising two Coulomb-coupled moiré lattices---offer a promising system for investigating strongly correlated dipolar excitons (composite bosons) with electrical control \cite{zhang_su4_2021}.
Thus far, their implementation has been hindered by the relative misalignment and incommensurability of the two \moire patterns \cite{zeng_exciton_2023}.
Here we report a dual \moire system with perfect translational and rotational alignment, achieved by utilizing twisted hexagonal boron nitride (hBN) bilayer to both generate an electrostatic \moire potential and separate MoSe$_{2}$ and WSe$_{2}$ monolayers.
We observe strongly correlated electron phases driven by intralayer interactions and identify interlayer Rydberg trions, which become trapped in the presence of the Mott insulating state.
Importantly, our platform is electrostatically programmable, allowing the realization of different lattice symmetries with either repulsive or attractive interlayer interactions.
In particular, we implement the latter scenario by optically injecting charges, which form a dipolar excitonic phase.
Our results establish a versatile platform for the exploration and manipulation of exotic and topological bosonic quantum many-body phases.
\end{abstract}

\maketitle

\section*{Main}

\begin{figure*}[t]
\includegraphics[width=1\textwidth]{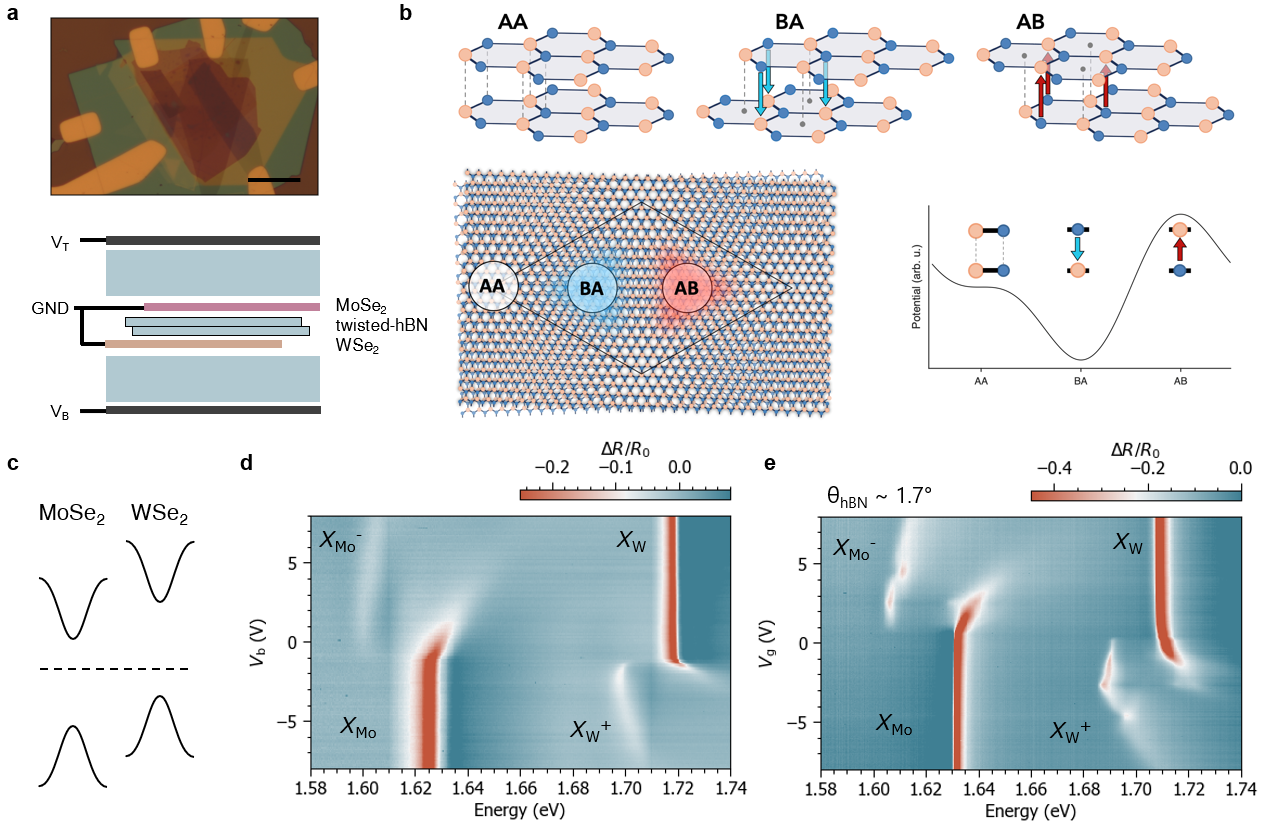}
\caption{\label{fig:device}\textbf{Device configuration and optical response of the Coulomb-coupled dual moiré system}
\newline
\textbf{a,} Optical micrograph (top) and schematic (bottom) of the Coulomb-coupled dual moiré system. A twisted hBN bilayer separates the WSe$_{2}$ and MoSe$_{2}$ monolayer. 
The system is symmetrically encapsulated by $\sim 45$ nm hBN layers and dual-gated using few-layer graphene flakes.
The scale bar is 20 $\mu$m.
\textbf{b,} Emergence of a periodic electrostatic potential at the interface of marginally twisted hBN bilayer.
Charge transfer at high symmetry points (AA, BA, AB) in the \moire unit cell (black diamond) gives rise to local electric polarization shown by blue and red arrows or shading, creating a periodic potential for charges with modulation $\Delta V\mathrm{_{e}}$. 
\textbf{c,} Schematic of the band alignment of the type-II heterojunction.
For a small electric field across the junction, electrons are injected into the MoSe$_{2}$ layer and holes into the WSe$_{2}$ layer.
\textbf{d,} Gate-dependent reflection contrast from a WSe$_{2}$/MoSe$_{2}$ atomic double-layer system separated by a natural (non-twisted) bilayer hBN, serving as a control experiment.
\textbf{e,} Gate-dependent reflection contrast from the Coulomb-coupled dual \moire system, showing the emergence of cusps at specific gate voltages corresponding to integer and fractional fillings of the \moire superlattice.
}
\end{figure*}

Gases of long-lived dipolar excitons confined to a periodic lattice constitute a fertile playground for studying and controlling coherent quantum many-boson phenomena \cite{baranov_condensed_2012, eisenstein_boseeinstein_2004}. 
In spaced double-layer systems, electrons and holes reside in separated adjacent layers that suppresses interlayer tunneling while maintaining strong Coulomb coupling. 
As a result, long-lived dipolar excitons form spontaneously and can be electrically injected near equilibrium \cite{eisenstein_boseeinstein_2004}.
In particular, such systems can be readily realized using two-dimensional (2D) materials by separating two transition metal dichalcogenides (TMD) or metal layers using a thin hBN spacer \cite{fogler_high-temperature_2014, ma_strongly_2021, jauregui_electrical_2019, wang_evidence_2019, liu_crossover_2022}.
However, the very spacer that confers this system with a long lifetime and electric control also suppresses the formation of a \moire lattice between the two layers.
Conversely, unspaced \moire bilayers exhibit a strong periodic \moire potential, which generates flat bands and enhances many-body interactions \cite{tang_simulation_2020, regan_mott_2020, xu_correlated_2020, wang_correlated_2020, shimazaki_strongly_2020}, but the short lifetime and the off-resonant optical excitation of interlayer excitons quench bosonic correlations and impose significant excess energy, driving their population far from thermal equilibrium \cite{wilson_excitons_2021}.

Attempts to introduce a lattice for long-lived dipolar excitons relied on coupling a \moire heterobilayer to a nearby \moire-free layer, with one charge of the dipolar exciton residing in each layer.
In this way, dipolar excitonic insulators \cite{zhang_correlated_2022, gu_dipolar_2022} and Bose-Fermi mixtures \cite{ben_mhenni_gate-tunable_2024} have been realized.
However, these systems do not realize the strong correlation regime for dipolar excitons, since the charge in the \moire-free layer resides in a dispersive band.
Another fundamental limitation is the maximum dipolar exciton density, which can only reach a fraction of the \moire density \cite{zhang_correlated_2022, ben_mhenni_gate-tunable_2024}.
A significant enhancement of correlations has been achieved by coupling two \moire heterobilayers together.
Here, both charges forming the dipolar excitons are hosted in flat bands \cite{zeng_exciton_2023}.
Though the constituent charges each experiences a periodic triangular potential, the dipolar exciton they form evolves in a random aperiodic potential that cannot be engineered, nor can it even be characterized.
This originates from the incommensurability of the two \moire lattices and the stochastic relative rotational and positional alignment.
Thus far, the challenge of engineering long-lived dipolar excitons that interact strongly in a periodic lattice has not been overcome.
The study of the strongly interacting exciton Hubbard models, as well as the exploration and manipulation of a plethora of exotic collective bosonic phenomena such as exciton crystals, superfluids, supersolids, and topological exciton structures, hinges on a physical implementation of this platform \cite{zhang_su4_2021, zeng_layer_2022, zhang_doping_2022, xie_long-lived_2024, zhang_engineering_2025}.

In this work, we experimentally realize a separated MoSe$_{2}$/WSe$_{2}$ system where long-lived dipolar excitons experience a strong periodic potential.
Charges in both TMD monolayers reside in \moire potentials with identical periodicities and perfect relative alignment.
We observe strongly correlated fermionic states within each layer, and strong interlayer correlations that manifest via the emergence of interlayer Rydberg trions and of a dipolar excitonic phase.

\subsection*{Inducing an electrostatic \moire superlattice via twisted hBN bilayers}

The observations are made possible by the dual-gated atomic double-layer system depicted in Fig.~\ref{fig:device}a.
The core of the device consists of MoSe$_{2}$ and WSe$_{2}$ monolayers, which are separated by a twisted hBN bilayer spacer (two hBN monolayers stacked on top of each other with a relative twist). This spacer fulfills different essential functions.
It separates the charges in the two TMD layers, while it simultaneously preserves strong intralayer Coulomb coupling. 
These ingredients allow for the stabilization of long-lived dipolar excitons.
Furthermore, it generates an electrostatic \moire potential \cite{zhao_universal_2021, kim_electrostatic_2024, wang_moire_2025} that permeates into both adjacent TMD monolayers.

Figure~\ref{fig:device}b elucidates the origin of the electrostatic \moire potential in twisted hBN bilayers.
Electric dipoles form due to charge transfer between neighbouring boron (B) and nitrogen (N) atoms across adjacent hBN layers.
When two hBN monolayers are twisted by a small angle (close to $0^{\circ}$), the resulting variation in local interlayer stacking registry across the moiré unit cell leads to a spatial modulation of the electric polarization \cite{li_binary_2017, zhao_universal_2021}.
This modulation becomes clear when examining three high-symmetry points, which occur within the \moire unit cell and approximate AA, BA, and AB stacking, respectively.
In AA stacking, identical atoms align vertically (B over B, N over N), and therefore, no electric dipoles form.
In BA stacking, N atoms in the upper layer lie directly above B atoms in the lower layer, resulting in downward electric dipoles. 
No opposing upward electric dipoles are produced since the B(N) atoms in the upper(lower) layer are vertically aligned with the hollow site of the adjacent layer \cite{constantinescu_stacking_2013}.
Therefore, BA stacking leads to \textit{downward} polarization.
Conversely, AB stacking results in the reversed configuration and thus leads to \textit{upward} polarization.

In essence, a triangular electrostatic \moire potential emerges at the interface of the twisted hBN layers with an amplitude that decays as $\mathrm{exp}(-d\frac{4\pi}{\sqrt{3}a\mathrm{_{moire}}})$ with the distance d \cite{zhao_universal_2021}.
Importantly, the MoSe$_{2}$ and the WSe$_{2}$ monolayers both share an interface with the twisted hBN bilayer, and therefore, they are each subject to a strong triangular \moire potential with identical \moire periodicity.
Specifically, the twist angle in the main device is $\delta_{\text{t}} \sim  1.7^{\circ}$, which corresponds to a \moire periodicity $a_{\text{moire}} \sim 8.5$ nm ($a_{\text{moire}} = a_{\text{hBN}}/\delta_{\text{t}}$ with $a_{\text{hBN}} \sim 2.5$ $\text{\AA}$ the lattice constant of hBN) \cite{zhao_universal_2021, pease_crystal_1950} and \moire density $n_{\text{moire}} \sim 1.6 \times 10^{12} \text{cm}^{-2}$.

Additionally, the device is dual-gated, allwing the independent tuning of the electric field and the carrier density of the dual \moire system, and therefore, the tuning of the filling factors $(\nu_\text{Mo} = n_\text{Mo}/n_\text{moire}, \nu_\text{W} = n_\text{W}/n_\text{moire})$ of the MoSe$_{2}$ and the WSe$_{2}$ \moire superlattices, with $n_\text{Mo}$ and $n_\text{W}$ the carrier densites of the respective layers.
The bottom (top) gate voltage is denoted by $V_{\text{b}}$ ($V_{\text{t}}$), while we use $V_{\text{g}}$ when the same voltage is applied to both bottom and top gates.
Particularly, for small electric fields across the atomic double-layer system, the type-II band alignment of the TMD layers is preserved (Fig.~\ref{fig:device}c), such that electrostatically injected electrons are hosted in the MoSe$_{2}$ layer, and holes in the WSe$_{2}$.

We start by presenting results obtained from a control device without a superlattice potential, which will allow us to subsequently isolate the effects of the twisted hBN bilayer.
Specifically, this device features an identical architecture to the main device except for the spacer, which is an as-exfoliated (non-twisted) bilayer hBN instead.
Figure ~\ref{fig:device}d shows gate-dependent reflection contrast spectra ($\Delta R / R_{0}$) of the control device.
Consistent with the type II band alignment, positive $V\mathrm{_{b}}$ injects electrons in the MoSe$_{2}$ layer.
This manifests via the presence of the negative MoSe$_{2}$ trion ($X_\text{Mo}^{-}$) and the neutral  WSe$_{2}$ exciton ($X_\text{W}$). 
Near $V\mathrm{_{b}}\sim -1$ V, both layers are charge-neutral, which is clear from the coexistence of the neutral exciton of  MoSe$_{2}$ ($X_\text{Mo}$) and $X_\text{W}$.
Conversely, negative $V\mathrm{_{b}}$ injects holes in the WSe$_{2}$ layer.
Here, the optical response is dominated by $X_\text{Mo}$ and the positive trion of WSe$_{2}$ ($X_\text{W}^{+}$).

Figure~\ref{fig:device}e shows gate-dependent reflection contrast spectra of the twisted hBN bilayer device.
Its general electrostatic behavior is similar to that of the control device (Fig~\ref{fig:device}d), showing layer-dependent electron/hole doping.
However, in stark contrast to the control device, the trions and the polaronic branches of the neutral excitons display prominent cusps at equally spaced discrete voltages.
The cusps are also evident in the spectra of monolayer MoSe$_{2}$ and WSe$_{2}$ outside the atomic double-layer system. (See Extended Data Fig.~\ref{figext:monolayers})
This striking behavior, which is enabled by the twisted hBN spacer, is reminiscent of the cusping of \moire excitons in semiconducting \moire bilayers caused by the emergence of quantum phases driven by strong correlations \cite{tang_simulation_2020, xu_correlated_2020, ben_mhenni_gate-tunable_2024}.
The resemblance suggests that strongly correlated electron phases emerge at special fillings of the electrostatic \moire superlattices generated by the twisted-hBN interface at the TMD layers. 
However, the optical response of each TMD layer is still defined by their neutral and charged excitons, contrary to the emergence of \moire excitons in semiconducting \moire bilayers \cite{tang_simulation_2020, regan_mott_2020, xu_correlated_2020}.
In fact, while the electrostatic \moire superlattice causes a potential modulation large enough for charge localization (on the order of $100$ meV) \cite{zhao_universal_2021}, its effect on excitons is much weaker. 
The underlying reason is the in-plane polarizability of the intralayer excitons. 
Based on an exciton polarizability of $6.5$ eV nm$^{2}$ V$^{-2}$ \cite{cavalcante_stark_2018} and estimating the maximum in-plane field to be on the order of 10$^{-2}$ V/nm, the $1s$ exciton would be subject to a potential modulation on the order of $1$ meV arising from the DC Stark effect.
This energy scale is much smaller than the exciton binding energy, explaining why the exciton spectra of the twisted hBN device do not exhibit \moire excitons.

\subsection*{Mott state and trapping of Rydberg interlayer trions}

\begin{figure*}[t]
\includegraphics[width=1\textwidth]{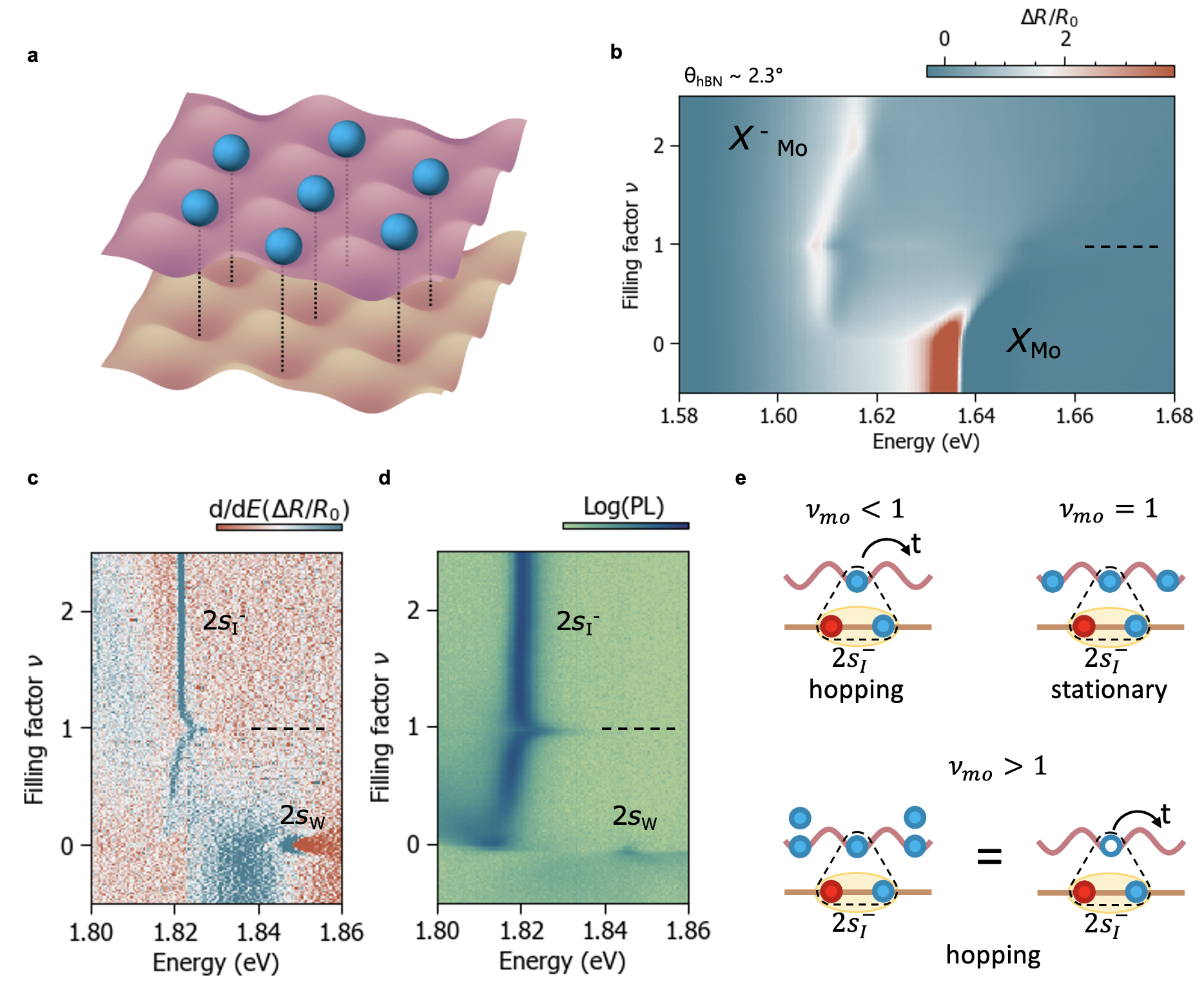}
\caption{\label{fig:mott}\textbf{Signatures of the Mott insulating phase and trapping of 2$s$ interlayer trions.}
\newline
\textbf{a,} Schematic of an electron Mott insulating phase realized on the Coulomb-coupled dual moiré system at $(\nu _\text{Mo}, \nu _\text{W}) = (1,0)$.
\textbf{b,} Gate-dependent reflection contrast sweep highlighting $X_\text{Mo}$ and $X_\text{Mo}^{-}$.
\textbf{c,} First derivative with respect to energy of the gate-dependent reflection contrast sweep, highlighting the Ryberg excitons.
\textbf{d,} Gate-dependent PL sweep highlighting the Ryberg excitons.
\textbf{e,} Nature of $2s_\text{I}^\text{-}$ and mechanism of its trapping at the Mott phase.
Note: The data is from Device II.
}
\end{figure*}

We proceed by studying a second device (Device II) with a slightly larger twist angle $\delta_{\text{t}} \sim  2.3^{\circ}$, which realizes a \moire periodicity $a_{\text{moire}} \sim 6.2$ nm and \moire density $n_{\text{moire}} \sim 3 \times 10^{12} \text{cm}^{-2}$.
The configuration $(\nu _\text{Mo}, \nu _\text{W}) = (1,0)$, where the electron Mott insulator (Fig.~\ref{fig:mott}a) would emerge is of particular interest.
Consequently, we initially focus on the excitons of MoSe$_{2}$.
Figure~\ref{fig:mott}b shows gate-dependent reflection contrast spectra as a function of filling factor measured on Device II and plotted over the spectral and doping region of interest.
Both $X_\text{Mo}^{-}$ and the polaronic branch of $X_\text{Mo}$ show clear cusps at $(\nu _\text{Mo}, \nu _\text{W}) = (1,0)$, and match the observations made in our main device (Fig.~\ref{fig:device}e).
The abrupt redshift of both exciton species and the increase in oscillator strength of $X_\text{Mo}^{-}$ are consistent with the emergence of a Mott insulating state \cite{tang_simulation_2020, kiper_confined_2025}.
Additionally, a detailed analysis of $X_\text{Mo}^{-}$ and $X_\text{W}^{+}$ reveals signatures of electron and hole generalized Wigner crystals (GWCs) at fractional fillings, Mott and band insulators at integer fillings (See Extended data Fig.~\ref{figext:analysis_device1}), which is in agreement with recent observations in MoSe$_{2}$ \cite{kiper_confined_2025}.

To further support our findings, we turn to study the Rydberg excitons in the WSe$_{2}$ layer.
In WSe$_{2}$, the $2s$ exciton has a Bohr radius of about $6$ nm \cite{stier_magnetooptics_2018}.
This makes it extremely sensitive to screening \cite{ben_mhenni_breakdown_2025} from mobile charges in the MoSe$_{2}$, which is situated at a distance less than $2$ nm.
Previous studies have used the $2s$ exciton in a sensor layer to detect the presence of incompressible correlated states in a nearby \moire layer, wherein a sudden loss of screening in phases such as the Fermi-Hubbard states \cite{xu_correlated_2020} or the dipolar excitonic insulators \cite{zhang_correlated_2022, ben_mhenni_gate-tunable_2024} leads to the reemergence of the $2s$ exciton. 
This behavior of Rydberg excitons has also been explained by a mixing of $2s$ and $2p$ states mediated by an electron crystal \cite{kim_excitons_2025}.

Figure ~\ref{fig:mott}c shows the first derivative of the gate-dependent reflection contrast sweep extracted from the same data as Fig.~\ref{fig:mott}b, but now highlighting the energy range of WSe$_{2}$ Rydberg excitons.
At charge neutrality, we observe the $2s_\text{W}$ exciton, which blueshifts and fades away upon injecting electrons in the proximal MoSe$_{2}$.
Against our expectations, the $2s_\text{W}$ exciton does not reemerge at any electron doping of the MoSe$_{2}$, including the Mott insulator at $(\nu _\text{Mo}, \nu _\text{W}) = (1,0)$. 
Instead, the spectra reveal a different resonance, labeled $\xin$, which emerges upon electron doping of the MoSe$_{2}$ while the WSe$_{2}$ is charge-neutral.
Furthermore, $\xin$ exhibits a pronounced cusp at $(\nu _\text{Mo}, \nu _\text{W}) = (1,0)$, concurrent with $X_\text{Mo}^{-}$ and the polaronic branch of $X_\text{Mo}$.
We corroborate the observation of $\xin$ via PL, where it exhibits a consistent behavior (Fig.~\ref{fig:mott}d).
To interpret the response of $\xin$ to the electronic state at $(\nu _\text{Mo}, \nu _\text{W}) = (1,0)$, it is necessary to first understand the origin of the resonance. 

We first note that $\xin$ is visible in the control device with the non-twisted hBN bilayer (See Extended Data Fig.~\ref{figext:rydberg}).
There, it exhibits similar behavior and characteristics to the twisted hBN bilayer device, except for the absence of any cusping behavior at discrete electron densities. 
Therefore, we conclude that the \moire potential is not required for the formation of $\xin$ itself.
Energetically, $\xin$ is redshifted by $\sim 25$ meV relative to the $2s_\text{W}$ exciton.
This value is close to typical trion binding energies in TMD monolayers. 
In MoSe$_{2}$ the trion binding energy of $X^{-}$/$X^{+}$ ($2s^{-}$/$2s^{+}$) is $\sim 23$ meV ($\sim 16-22$ meV).
While in WSe$_{2}$, the trion binding energy of $X^{-}_{\text{S,T}}$/$X^{+}$ ($2s^{-}$/$2s^{+}$) is $\sim 27-35$ meV ($\sim 17$ meV) \cite{liu_exciton-polaron_2021}.
Moreover, $\xin$ is anticorrelated with $2s_\text{W}$, which reinforces the hypothesis that the former is a charged quasiparticle derived from the latter.
However, since the WSe$_{2}$ is charge-neutral, we exclude the intralayer $2s^{-}_\text{W}$.
In this case, the charge can only originate from the MoSe$_{2}$ layer, which is electron-doped, since the system is in the configuration $(\nu _\text{Mo}>0,  \nu _\text{W}=0)$.
Consequently, we identify $\xin$ as being a Rydberg interlayer trion, emerging when a $2s_\text{W}$ residing in the WSe$_{2}$ binds to an electron in the MoSe$_{2}$ layer via charge-dipole interactions.
We observe an analogous behavior for the $\xip$, in which the exciton part of the wavefunction resides in the MoSe$_{2}$, and the charge bound to it is a hole in the WSe$_{2}$ (See Extended Data Fig.~\ref{figext:rydberg}).

Based on this interpretation, we consider the effect of a correlated insulator on $\xin$ and how it causes it to blueshift.
In the regime $(\nu _\text{Mo}>0,  \nu _\text{W}=0)$, the electrons in MoSe$_{2}$ become localized upon emergence of an incompressible phase.
Since the extra electron---and not the excitonic part---of $\xin$ resides in the MoSe$_{2}$, there are two distinct scenarios for the quasiparticle when the electron becomes localized.

If the kinetic energy of $2s_\text{W}$ is larger than its binding energy to the electron in the MoSe$_{2}$, $\xin$ would dissociate into its constituting components when the charges become localized.
In this scenario, $\xin$ would disappear and $2s_\text{W}$ would reemerge at the Mott phase $(\nu _\text{Mo}=1,  \nu _\text{W}=0)$ close in energy its initial energy at $(\nu _\text{Mo}=0,  \nu _\text{W}=0)$ \cite{xu_correlated_2020}.
This scenario is unlikely since at $(\nu _\text{Mo}=0,  \nu _\text{W}=0)$, $\xin$ blueshifts only by a few meV without disappearing, and  $2s_\text{W}$ does not reappear.

The other possibility is that $\xin$ remains bound when incompressible phases emerge, as a consequence of its relatively large binding energy of $\sim 25$ meV relative to $2s_\text{W}$.
This, in turn, leads to two additional eventualities, depicted in Fig.~\ref{fig:mott}e.
Starting from a hopping behaviour, either the $\xin$ becomes trapped in a \moire superlattice site upon emergence of an incompressible phase, if this phase is thermodynamically stable enough to quench the kinetic energy of the quasiparticle.
In contrast, if the opposite condition is true, $\xin$ continues its hopping through the system.
In the experiments (Fig.~\ref{fig:mott}c,e), we observe a cusp at the Mott insulator $(\nu _\text{Mo}=1,  \nu _\text{W}=0)$, but do not observe any other cusp in $\xin$, although a cusp is visible in $X_\text{Mo}^{-}$ at the band insulator $(\nu _\text{Mo}=2,  \nu _\text{W}=0)$.
Among the phases of the extended Fermi-Hubbard model on a triangular lattice, the Mott insulator is by far the most thermodynamically stable state \cite{xu_correlated_2020}.
In this picture, $\xin$ is trapped in the strong Mott phase, but its kinetic energy outcompetes localization for other phases, such as the weaker GWCs and the band insulator.
The interaction of $\xin$ with incompressible phases, which is likely governed by a hopping/trapping mechanism, is distinct from other optical probes, such as those relying on dielectric screening or Umklapp spectroscopy. Thereby, this could provide access to hidden microscopic parameters of the system. 

\subsection*{Melting of the Mott state}
\begin{figure*}
\includegraphics[width=\textwidth]{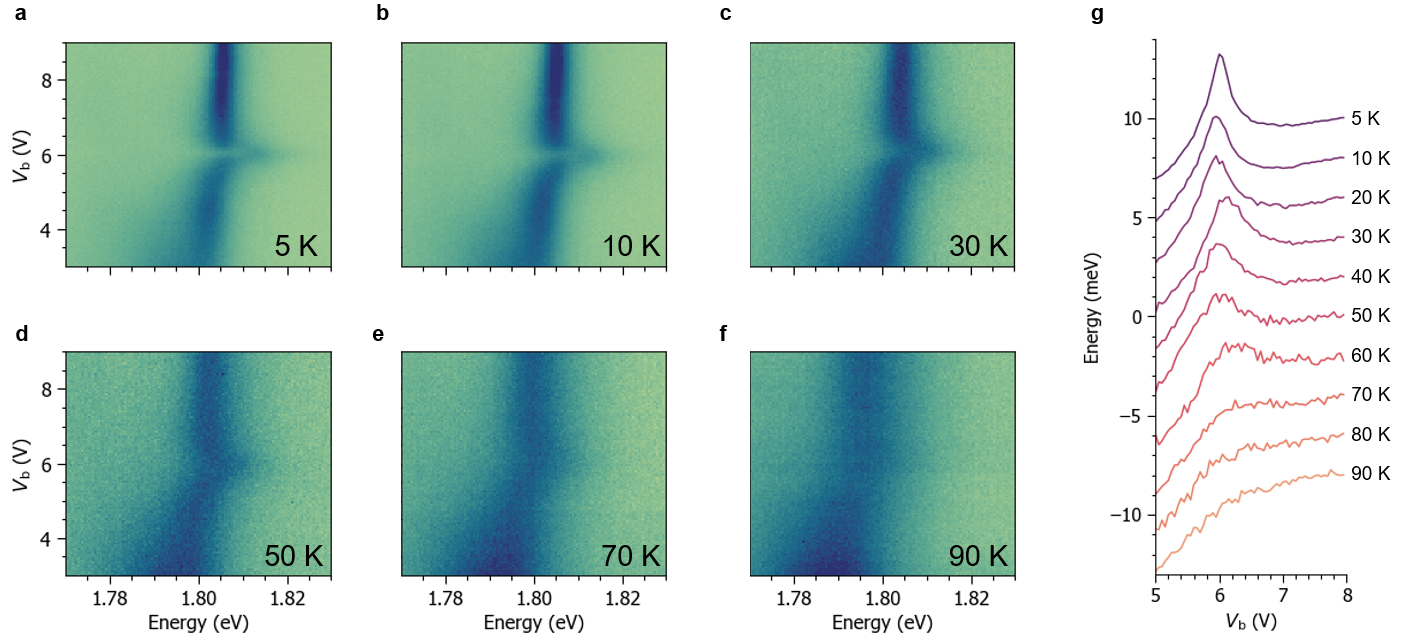}
\caption{\label{fig:melting}\textbf{Melting of the electron Mott insulating phase.}
\newline
\textbf{a-f,} Gate-dependent PL sweeps showing the $2s_\text{I}^\text{-}$ at $5$ K (\textbf{a}), $10$ K (\textbf{b}), $30$ K (\textbf{c}), $50$ K (\textbf{d}), $70$ K (\textbf{e}), and $90$ K (\textbf{f}).
\textbf{g,} Energy of the $2s_\text{I}^\text{-}$ extracted via fitting of the gate-dependent PL sweeps at the different temperatures.
The $2s_\text{I}^\text{-}$ trapping occurs up to a temperature of $60$ K, and thus, the Mott insulating phase is stable up to at least $60$ K.
Note: The data is from the main device.
}
\end{figure*}

We continue our study of the Mott insulator by investigating its melting temperature via monitoring the cusp of $\xin$.
Figure~\ref{fig:melting}a-f shows the gate-dependent PL of the main device at different temperatures in the range $5-90$ K.
As the temperature increases, the cusp corresponding to the Mott insulator gradually wanes and completely disappears between $50-70$ K. 
To monitor the blueshifting at the cusp more precisely, we extract the $\xin$ central energy via fitting of a Lorentzian lineshape on a more granular data set (Fig.~\ref{fig:melting}g).
The blueshifting behavior of $\xin$ at $(\nu _\text{Mo}=1,  \nu _\text{W}=0)$ weakens gradually with increasing temperature, which is indicated by the gradual decrease of the cusp prominence in Fig.~\ref{fig:melting}g. 
By $70$ K, the cusp is essentially absent.
At this temperature, $\xin$ evolves continuously with $\nu _\text{Mo}$ akin to the control sample.
The disappearance of the $\xin$ cusp and the similarity of its behavior to the control sample, which lacks a superlattice potential, point towards the melting of a correlated electron state into a population of mobile charges. 
In that case, the trapping of $\xin$ cannot take place anymore.
Based on the melting temperature of around $60-70$ K, we estimate the correlation gap of the state to be around $6$ meV, notably smaller than in moire heterobilayers with similar periodicity.
We note, however, that this estimation corresponds to the temperature at which  $\xin$, rather than a bare electron in the MoSe2, is no longer localized by the Mott insulator.
Therefore, the actual melting temperature of the Mott state could be slightly higher.
Finally, the melting temperature of the Mott state in Device II is around $40-50$ K (Extended Data Fig.~\ref{figext:melting_device2}). 
This slightly lower value can be explained by Device II's larger twist angle and correspondingly larger charge kinetic energy. 

\subsection*{Programmable geometry of the dual \moire system and optical injection of a dipolar excitonic phase}

\begin{figure*}
\includegraphics[width=\textwidth]{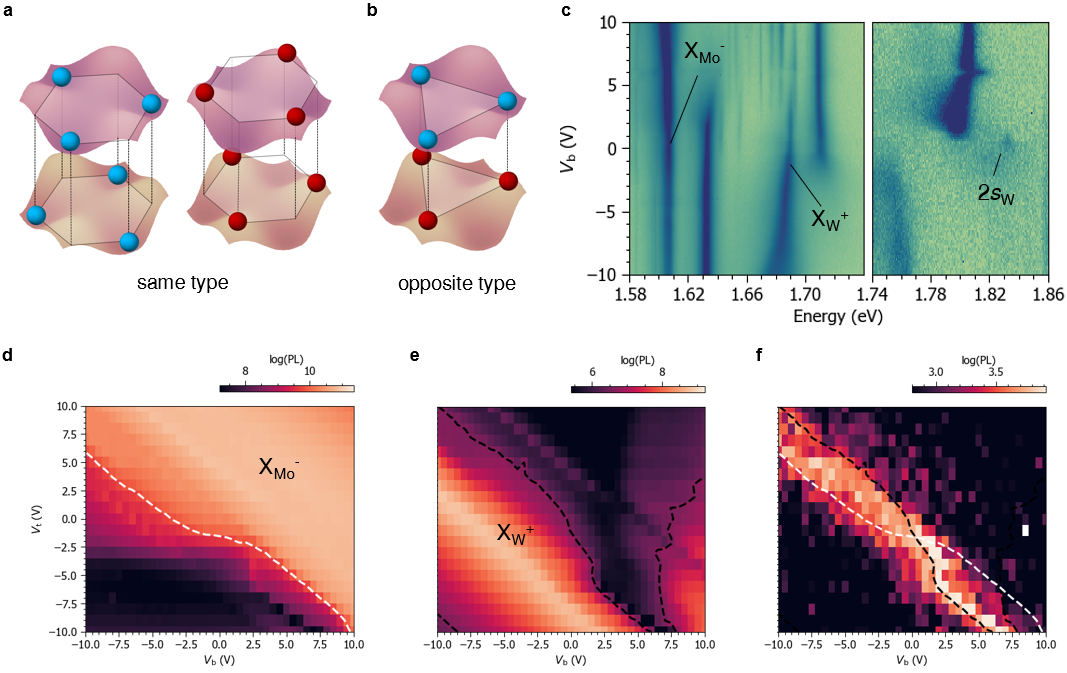}
\caption{\label{fig:geometry}\textbf{Programmable geometry of the dual \moire system and optical injection of dipolar excitons.}
\newline
\textbf{a,} Sketch of the dual \moire system's geometry when charges of the same type---either electrons (right) or holes (left)---are simultaneously injected in both layers. Here, the illustrative cases $(\nu _\text{Mo}, \nu _\text{W}) = (1,1)$ and $(-1,-1)$ are shown.
\textbf{b,} Sketch of the dual \moire system's geometry when electrons are injected in the MoSe$_{2}$ and holes in WSe$_{2}$ layer simultaneously. Here, the illustrative case of $(\nu _\text{Mo}, \nu _\text{W}) = (1,-1)$ is shown.
\textbf{c,} Gate-dependent PL sweep showing the coexistence of $X_\text{Mo}^{-}$ in MoSe$_{2}$, the $X_\text{W}^{+}$ in WSe$_{2}$, and $2s_\text{W}$.
\textbf{d-f,} Electrostatic phase diagram of $X_\text{Mo}^{-}$ (\textbf{d}), of $X_\text{W}^{+}$ (\textbf{e}), and of $2s_\text{W}$ (\textbf{f}). The PL spectra are integrated over the range $1.59$ to $1.62$ eV, $1.687$ to $1.693$ eV, and $1.835$ to $1.845$ eV, respectively. 
The $2s_\text{W}$ persists over an area of the electrostatic phase diagram where electrons (in the MoSe$_{2}$) with holes (in the WSe$_{2}$) coexist in the dual \moire system. 
Note: The data is from the main device.
}
\end{figure*}

Thus far, we have investigated the dual \moire system in a regime where the charge density of one layer is selectively tuned, while the other layer remains charge-neutral.
Under these conditions, strongly correlated phases of electrons on a lattice emerge. 
In addition, interlayer excitonic states $\xin$ and $\xip$ arise under optical pumping and interact with the Mott insulating phase.
For these phases, the relative rotational and translational alignment of the constituent superlattices does not play a significant role.

Conversely, the system accesses a different phase space when we electrostatically inject charges in both layers simultaneously.
The charges across the layers experience strong intralayer correlations since the separation of the two TMD layers is significantly smaller than the \moire period, and concomitantly, the interlayer Coulomb interactions are comparable in magnitude to the intralayer ones.
In particular, this makes the formation of composite quasiparticles possible.
These include dipolar excitons, which can realize bosonic collective phenomena such as excitonic insulators, exciton density waves, and exciton superfluidity, and even Bose-Fermi mixtures when coexisting with fermions.

In this interlayer regime, the relative alignment of the constituent superlattices becomes of fundamental importance, since it governs the effective lattice geometry, contrasting sharply with the intralayer scenario. 
In particular, random or aperiodic misalignment, for example in the case of two separate \moire heterobilayers stacked on top of each other, drastically alters the quantum many-body phase diagram and makes it hard to reproducibly access, study, and interpret exotic phases of interest \cite{zeng_layer_2022}.
In the following, we will continue to elaborate on the alignment and the programmable geometry of the dual \moire system, which are unique to our approach.
Subsequently, the emergence of a dipolar exciton phase will be discussed and explored throughout the electrostatic phase diagram.

The constituent superlattices of our dual \moire system stem from the same potential source, guaranteeing the rotational and translational alignment between charge lattices in the two layers in a straightforward fashion.
Since the potential originates from electric polarization at the twisted bilayer hBN, it is antisymmetric across the mirror plane between the layers.
For instance, locales with upward electric polarization correspond to the potential minima for electrons in the upper layer (in our device, the MoSe$_{2}$ layer), and at the same time, for holes in the lower layer (in our device, the WSe$_{2}$ layer).
Even inhomogeneities and distortions in the source \moire superlattice will be propagated in a consistent fashion across the two TMD layers.

Concretely, the system geometry depends on whether the layers host charges of the same or opposite sign.
When considering charges of the same type injected into both layers (Fig.~\ref{fig:geometry}a), the potential minima in the two layers are not vertically stacked, but are effectively rotated by $180^{\circ}$ degrees, resulting in a hexagonal superlattice unit cell with broken inversion symmetry as alternating corners in the hexagonal supercell are populated by charges in opposite layers.
This regime is accessible by applying large electric fields using the top and bottom gates (Extended Data Fig.~\ref{figext:same_type}).
When the charges of each layer have opposite signs (Fig.~\ref{fig:geometry}b), then the potential minima for electrons in MoSe$_{2}$ sit directly on top of the minima for holes in WSe$_{2}$.
Here, the resulting superlattice structure has a triangular unit cell with threefold rotational symmetry. 

Having established the perfect alignment---both rotational and positional, and the configurable geometry of our dual \moire system, which are key characteristics when charges coexist in both layers, our attention now moves to studying its optical response under such conditions.
Figure.~\ref{fig:geometry}c shows gate-dependent ($V\mathrm{_{b}}$) PL recorded from the main device. 
The coexistence of $X_\text{Mo}^{-}$ and $X_\text{W}^{+}$ over a wide range of $V\mathrm{_{b}}$ indicates the simultaneous presence of opposite carriers in the system---electrons in MoSe$_{2}$ and holes in WSe$_{2}$.
This charge configuration arises without an applied bias voltage between the two TMD layers. 
Rather, the charging results from optical injection of opposite charges in the respective layers due to the type-II band alignment and weak but nonzero interlayer tunneling.
This is consistent with a recent report \cite{tugen_optical_2025}.
Here, $X_\text{Mo}^{-}$ persists over a wider voltage range, suggesting a more efficient electron injection in the MoSe$_{2}$, likely due to the dielectric stack being specifically optimized for that spectral range.
Surprisingly, the $2s_\text{W}$ emerges within the range over which $X_\text{Mo}^{-}$ and $X_\text{W}^{+}$ coexist.
In the absence of quantum many-body effects, not only would the $2s_\text{W}$ disappear as soon as sufficient holes are injected into WSe$_{2}$ and the $X_\text{W}^{+}$ appears, but also as soon as electrons are injected into the proximal MoSe$_{2}$ and the $X_\text{Mo}^{-}$ appears.
Therefore, the observed emergence of $2s_\text{W}$, despite this charge configuration, indicates that the charges present in the system ineffectively screen the $2s$ excitons.
This phenomenon can be explained by the formation of dipolar excitons---composite bosons formed by electrons in MoSe$_{2}$ binding to holes in WSe$_{2}$.
Device II exhibits similar behavior, further corroborating these observations (See Extended Data Fig.~\ref{figext:dipolar}).

To delimit the dipolar excitonic phase, we track this behavior throughout the electrostatic phase diagram. 
We start by identifying the regions of electron (hole) doping of the MoSe$_{2}$ (WSe$_{2}$) layer, by monitoring $X_\text{Mo}^{-}$ ($X_\text{W}^{+}$).
Figures~\ref{fig:geometry}d,e show the PL electrostatic phase diagrams of $X_\text{Mo}^{-}$ and $X_\text{W}^{+}$, respectively.
We obtain them by integrating the PL signal over the spectral ranges $1.59$ to $1.62$ eV (Fig.~\ref{fig:geometry}d) and $1.687$ to $1.693$ eV (Fig.~\ref{fig:geometry}e), which correspond to the energy ranges over which $X_\text{Mo}^{-}$ and $X_\text{W}^{+}$ have an almost constant peak energy before they start redshifting at high electron and hole densities, respectively.
We determine the boundaries (white and black dashed lines) by thresholding the integrated PL intensity of the trions (See Methods).

Concurrently, we monitor the $2s_\text{W}$ signal by integrating the same PL data over the energy range $1.835$ to $1.845$ eV (Fig.~\ref{fig:geometry}f).
The PL of $2s_\text{W}$ exists over approximately the anti-diagonal of the electrostatic phase diagram, which corresponds to the electric field axis of the system.
The $2s_\text{W}$ persists over the subspace, where electron doping of MoSe$_{2}$ and hole doping of WSe$_{2}$ coexist (white dashed curve). 
In this region, the number of electrons and holes in their respective layers are approximately equal. 
Therefore, this constitutes the subspace over which the undoped dipolar excitonic phase should exist.
Remarkably, the dipolar phase only exists for positive electric fields along the direction MoSe$_{2}$-WSe$_{2}$, which corresponds to the polarity that closes the type-II band gap.
Effectively, the optical injection of dipolar excitons in the dual \moire system can be switched on and off using the applied electric field.
In addition, the broadening of this region along the electrostatic doping axis (positive diagonal of the electrostatic phase diagram) suggests a mixing of the dipolar phase with itinerant charges.

\subsection*{Outlook}

We have realized experimentally a Coulomb-coupled dual \moire system with a well-defined and reproducible geometry.
By employing a twisted hBN spacer to induce an electrostatic \moire potential, we achieve identical periodicities and perfect registry of the two constituent lattices.
Each layer hosts strong intralayer correlations---most notably, a Mott insulating state with a melting temperature near $70$ K---while strong coupling between the two \moire layers gives rise to interlayer Rydberg trion formation and to the emergence of a dipolar excitonic phase under non-resonant optical excitation.
Beyond its electrostatic programmability---where two distinct geometries can be implemented---the architecture of the system admits further tailoring through the choice of the TMD materials and the thickness of the spacer.
For example, if two layers of the same TMD are used, the layer degeneracy will give rise to a layer pseudospin degree of freedom due to the relatively weak (but nonzero) interlayer tunneling through bilayer hBN \cite{ma_strongly_2021}. 
Since the layer pseudospin couples to electric fields, the twisted hBN would impart a spatial texture to the pseudospin. This texture includes chiral centers with nonzero winding numbers, exactly analogous to twisted TMD homobilayer systems, which may give rise to topological flat bands.
Furthermore, quantum emitters could be readily implanted in the twisted hBN spacer \cite{carbone_quantifying_2025} to sense emergent quantum phases.
Such design flexibility opens a robust avenue for exploring exotic many-boson phenomena in a clean, reproducible setting.
An immediate extension to our work would be to use time-resolved reflection spectroscopy \cite{tugen_optical_2025} to probe the strongly correlated exciton Hubbard model.
By combining electrostatic biasing and fast optical injection, one could tune the exciton density and search for lifetime enhancement at fractional and integer fillings of the \moire lattice, a hallmark of emergent exciton crystals.
Finally, this platform is ideally suited for investigating exciton superfluidity and supersolidity \cite{zhang_su4_2021}.

\section*{\label{sec:Methods}Methods}
\subsection*{\label{sec:methods:sample}Sample preparation}
All TMD, graphite, and hBN flakes were mechanically exfoliated from bulk crystals on $70$ nm SiO$_{2}$ substrates. 
Homogeneous flakes were selected based on their optical contrast, shape, and cleanliness.
The device was assembled via the dry-transfer technique using polycarbonate (PC) films in two parts.
The hBN monolayer was cut into two using an atomic force microscope tip.
The picking up of flakes was done at $120 ^{\circ}$ C.
After assembly of the heterostructure on the stamp, the interfaces were cleaned by repeatedly contacting and picking up the heterostructure at $155 ^{\circ}$ C, which mechanically squeezes out trapped bubbles between the constituent layers.
The sample was subsequently submerged in chloroform ($30$ min) and IPA ($10$ min) to remove the polymer.
Based on a transfer matrix approach for plane wave propagation, the contrast of the dielectric stack is optimized by adding a further hBN flake on top of the heterostructure.
Finally, the respective layers are contacted using standard maskless optical lithography and subsequent electron beam evaporation of Cr/Au $5/100$ nm electrodes. 

\subsection*{\label{sec:methods:spectroscopy}Optical spectroscopy}
The optical measurements were performed in a close-cycle optical cryostat in reflection geometry (Attocube, attoDRY800).
For the reflection contrast measurements, white light from a  supercontinuum white light fiber laser was focused onto the sample using a $40$X objective with a numerical aperture of $
0.75$, yielding an excitation spot size of around $1$ $\mu$m. 
A pinhole was used as a spatial filter to obtain a diffraction-limited collection spot. The collected light was dispersed using a $500$ mm focal length spectrometer and detected on a back-illuminated CCD sensor array.
For the PL measurements, a $632.8$ nm helium-neon laser was used.
The gates were controlled using a source-measure unit with monitored leakage current.
The filling factors were calculated using the capacitor model. 
The hBN thicknesses were determined using atomic force microscopy, and a dielectric constant of $3.8$ was used for the hBN.
The cusps at the integer filling factors were used to extract an accurate twist angle and correct the model.
Unless otherwise specified, all measurements presented here were performed at $5$ K.

\section*{\label{sec:data_availability}Data availability}
The data sets generated and analyzed during the current study are available from the corresponding authors upon reasonable request.

\bibliography{references}

\begin{thebibliography}{43}%
\makeatletter
\providecommand \@ifxundefined [1]{%
 \@ifx{#1\undefined}
}%
\providecommand \@ifnum [1]{%
 \ifnum #1\expandafter \@firstoftwo
 \else \expandafter \@secondoftwo
 \fi
}%
\providecommand \@ifx [1]{%
 \ifx #1\expandafter \@firstoftwo
 \else \expandafter \@secondoftwo
 \fi
}%
\providecommand \natexlab [1]{#1}%
\providecommand \enquote  [1]{``#1''}%
\providecommand \bibnamefont  [1]{#1}%
\providecommand \bibfnamefont [1]{#1}%
\providecommand \citenamefont [1]{#1}%
\providecommand \href@noop [0]{\@secondoftwo}%
\providecommand \href [0]{\begingroup \@sanitize@url \@href}%
\providecommand \@href[1]{\@@startlink{#1}\@@href}%
\providecommand \@@href[1]{\endgroup#1\@@endlink}%
\providecommand \@sanitize@url [0]{\catcode `\\12\catcode `\$12\catcode `\&12\catcode `\#12\catcode `\^12\catcode `\_12\catcode `\%12\relax}%
\providecommand \@@startlink[1]{}%
\providecommand \@@endlink[0]{}%
\providecommand \url  [0]{\begingroup\@sanitize@url \@url }%
\providecommand \@url [1]{\endgroup\@href {#1}{\urlprefix }}%
\providecommand \urlprefix  [0]{URL }%
\providecommand \Eprint [0]{\href }%
\providecommand \doibase [0]{https://doi.org/}%
\providecommand \selectlanguage [0]{\@gobble}%
\providecommand \bibinfo  [0]{\@secondoftwo}%
\providecommand \bibfield  [0]{\@secondoftwo}%
\providecommand \translation [1]{[#1]}%
\providecommand \BibitemOpen [0]{}%
\providecommand \bibitemStop [0]{}%
\providecommand \bibitemNoStop [0]{.\EOS\space}%
\providecommand \EOS [0]{\spacefactor3000\relax}%
\providecommand \BibitemShut  [1]{\csname bibitem#1\endcsname}%
\let\auto@bib@innerbib\@empty
\bibitem [{\citenamefont {Baranov}\ \emph {et~al.}(2012)\citenamefont {Baranov}, \citenamefont {Dalmonte}, \citenamefont {Pupillo},\ and\ \citenamefont {Zoller}}]{baranov_condensed_2012}%
  \BibitemOpen
  \bibfield  {author} {\bibinfo {author} {\bibfnamefont {M.~A.}\ \bibnamefont {Baranov}}, \bibinfo {author} {\bibfnamefont {M.}~\bibnamefont {Dalmonte}}, \bibinfo {author} {\bibfnamefont {G.}~\bibnamefont {Pupillo}},\ and\ \bibinfo {author} {\bibfnamefont {P.}~\bibnamefont {Zoller}},\ }\bibfield  {title} {\bibinfo {title} {Condensed {{Matter Theory}} of {{Dipolar Quantum Gases}}},\ }\href {https://doi.org/10.1021/cr2003568} {\bibfield  {journal} {\bibinfo  {journal} {Chem. Rev.}\ }\textbf {\bibinfo {volume} {112}},\ \bibinfo {pages} {5012} (\bibinfo {year} {2012})}\BibitemShut {NoStop}%
\bibitem [{\citenamefont {Bloch}\ \emph {et~al.}(2012)\citenamefont {Bloch}, \citenamefont {Dalibard},\ and\ \citenamefont {Nascimb{\`e}ne}}]{bloch_quantum_2012}%
  \BibitemOpen
  \bibfield  {author} {\bibinfo {author} {\bibfnamefont {I.}~\bibnamefont {Bloch}}, \bibinfo {author} {\bibfnamefont {J.}~\bibnamefont {Dalibard}},\ and\ \bibinfo {author} {\bibfnamefont {S.}~\bibnamefont {Nascimb{\`e}ne}},\ }\bibfield  {title} {\bibinfo {title} {Quantum simulations with ultracold quantum gases},\ }\href {https://doi.org/10.1038/nphys2259} {\bibfield  {journal} {\bibinfo  {journal} {Nature Phys}\ }\textbf {\bibinfo {volume} {8}},\ \bibinfo {pages} {267} (\bibinfo {year} {2012})}\BibitemShut {NoStop}%
\bibitem [{\citenamefont {Landig}\ \emph {et~al.}(2016)\citenamefont {Landig}, \citenamefont {Hruby}, \citenamefont {Dogra}, \citenamefont {Landini}, \citenamefont {Mottl}, \citenamefont {Donner},\ and\ \citenamefont {Esslinger}}]{landig_quantum_2016}%
  \BibitemOpen
  \bibfield  {author} {\bibinfo {author} {\bibfnamefont {R.}~\bibnamefont {Landig}}, \bibinfo {author} {\bibfnamefont {L.}~\bibnamefont {Hruby}}, \bibinfo {author} {\bibfnamefont {N.}~\bibnamefont {Dogra}}, \bibinfo {author} {\bibfnamefont {M.}~\bibnamefont {Landini}}, \bibinfo {author} {\bibfnamefont {R.}~\bibnamefont {Mottl}}, \bibinfo {author} {\bibfnamefont {T.}~\bibnamefont {Donner}},\ and\ \bibinfo {author} {\bibfnamefont {T.}~\bibnamefont {Esslinger}},\ }\bibfield  {title} {\bibinfo {title} {Quantum phases from competing short- and long-range interactions in an optical lattice},\ }\href {https://doi.org/10.1038/nature17409} {\bibfield  {journal} {\bibinfo  {journal} {Nature}\ }\textbf {\bibinfo {volume} {532}},\ \bibinfo {pages} {476} (\bibinfo {year} {2016})}\BibitemShut {NoStop}%
\bibitem [{\citenamefont {Browaeys}\ and\ \citenamefont {Lahaye}(2020)}]{browaeys_many-body_2020}%
  \BibitemOpen
  \bibfield  {author} {\bibinfo {author} {\bibfnamefont {A.}~\bibnamefont {Browaeys}}\ and\ \bibinfo {author} {\bibfnamefont {T.}~\bibnamefont {Lahaye}},\ }\bibfield  {title} {\bibinfo {title} {Many-body physics with individually controlled {{Rydberg}} atoms},\ }\href {https://doi.org/10.1038/s41567-019-0733-z} {\bibfield  {journal} {\bibinfo  {journal} {Nat. Phys.}\ }\textbf {\bibinfo {volume} {16}},\ \bibinfo {pages} {132} (\bibinfo {year} {2020})}\BibitemShut {NoStop}%
\bibitem [{\citenamefont {Zhang}\ \emph {et~al.}(2022)\citenamefont {Zhang}, \citenamefont {Regan}, \citenamefont {Wang}, \citenamefont {Zhao}, \citenamefont {Wang}, \citenamefont {Sayyad}, \citenamefont {Yumigeta}, \citenamefont {Watanabe}, \citenamefont {Taniguchi}, \citenamefont {Tongay}, \citenamefont {Crommie}, \citenamefont {Zettl}, \citenamefont {Zaletel},\ and\ \citenamefont {Wang}}]{zhang_correlated_2022}%
  \BibitemOpen
  \bibfield  {author} {\bibinfo {author} {\bibfnamefont {Z.}~\bibnamefont {Zhang}}, \bibinfo {author} {\bibfnamefont {E.~C.}\ \bibnamefont {Regan}}, \bibinfo {author} {\bibfnamefont {D.}~\bibnamefont {Wang}}, \bibinfo {author} {\bibfnamefont {W.}~\bibnamefont {Zhao}}, \bibinfo {author} {\bibfnamefont {S.}~\bibnamefont {Wang}}, \bibinfo {author} {\bibfnamefont {M.}~\bibnamefont {Sayyad}}, \bibinfo {author} {\bibfnamefont {K.}~\bibnamefont {Yumigeta}}, \bibinfo {author} {\bibfnamefont {K.}~\bibnamefont {Watanabe}}, \bibinfo {author} {\bibfnamefont {T.}~\bibnamefont {Taniguchi}}, \bibinfo {author} {\bibfnamefont {S.}~\bibnamefont {Tongay}}, \bibinfo {author} {\bibfnamefont {M.}~\bibnamefont {Crommie}}, \bibinfo {author} {\bibfnamefont {A.}~\bibnamefont {Zettl}}, \bibinfo {author} {\bibfnamefont {M.~P.}\ \bibnamefont {Zaletel}},\ and\ \bibinfo {author} {\bibfnamefont {F.}~\bibnamefont {Wang}},\ }\bibfield  {title} {\bibinfo {title} {Correlated interlayer exciton insulator in heterostructures of monolayer {{WSe2}}
  and moir{\'e} {{WS2}}/{{WSe2}}},\ }\href {https://doi.org/10.1038/s41567-022-01702-z} {\bibfield  {journal} {\bibinfo  {journal} {Nat. Phys.}\ }\textbf {\bibinfo {volume} {18}},\ \bibinfo {pages} {1214} (\bibinfo {year} {2022})}\BibitemShut {NoStop}%
\bibitem [{\citenamefont {Gu}\ \emph {et~al.}(2022)\citenamefont {Gu}, \citenamefont {Ma}, \citenamefont {Liu}, \citenamefont {Watanabe}, \citenamefont {Taniguchi}, \citenamefont {Hone}, \citenamefont {Shan},\ and\ \citenamefont {Mak}}]{gu_dipolar_2022}%
  \BibitemOpen
  \bibfield  {author} {\bibinfo {author} {\bibfnamefont {J.}~\bibnamefont {Gu}}, \bibinfo {author} {\bibfnamefont {L.}~\bibnamefont {Ma}}, \bibinfo {author} {\bibfnamefont {S.}~\bibnamefont {Liu}}, \bibinfo {author} {\bibfnamefont {K.}~\bibnamefont {Watanabe}}, \bibinfo {author} {\bibfnamefont {T.}~\bibnamefont {Taniguchi}}, \bibinfo {author} {\bibfnamefont {J.~C.}\ \bibnamefont {Hone}}, \bibinfo {author} {\bibfnamefont {J.}~\bibnamefont {Shan}},\ and\ \bibinfo {author} {\bibfnamefont {K.~F.}\ \bibnamefont {Mak}},\ }\bibfield  {title} {\bibinfo {title} {Dipolar excitonic insulator in a moir{\'e} lattice},\ }\href {https://doi.org/10.1038/s41567-022-01532-z} {\bibfield  {journal} {\bibinfo  {journal} {Nat. Phys.}\ }\textbf {\bibinfo {volume} {18}},\ \bibinfo {pages} {395} (\bibinfo {year} {2022})}\BibitemShut {NoStop}%
\bibitem [{\citenamefont {Lagoin}\ \emph {et~al.}(2022{\natexlab{a}})\citenamefont {Lagoin}, \citenamefont {Suffit}, \citenamefont {Baldwin}, \citenamefont {Pfeiffer},\ and\ \citenamefont {Dubin}}]{lagoin_mott_2022}%
  \BibitemOpen
  \bibfield  {author} {\bibinfo {author} {\bibfnamefont {C.}~\bibnamefont {Lagoin}}, \bibinfo {author} {\bibfnamefont {S.}~\bibnamefont {Suffit}}, \bibinfo {author} {\bibfnamefont {K.}~\bibnamefont {Baldwin}}, \bibinfo {author} {\bibfnamefont {L.}~\bibnamefont {Pfeiffer}},\ and\ \bibinfo {author} {\bibfnamefont {F.}~\bibnamefont {Dubin}},\ }\bibfield  {title} {\bibinfo {title} {Mott insulator of strongly interacting two-dimensional semiconductor excitons},\ }\href {https://doi.org/10.1038/s41567-021-01440-8} {\bibfield  {journal} {\bibinfo  {journal} {Nat. Phys.}\ }\textbf {\bibinfo {volume} {18}},\ \bibinfo {pages} {149} (\bibinfo {year} {2022}{\natexlab{a}})}\BibitemShut {NoStop}%
\bibitem [{\citenamefont {Lagoin}\ \emph {et~al.}(2022{\natexlab{b}})\citenamefont {Lagoin}, \citenamefont {Bhattacharya}, \citenamefont {Grass}, \citenamefont {Chhajlany}, \citenamefont {Salamon}, \citenamefont {Baldwin}, \citenamefont {Pfeiffer}, \citenamefont {Lewenstein}, \citenamefont {Holzmann},\ and\ \citenamefont {Dubin}}]{lagoin_extended_2022}%
  \BibitemOpen
  \bibfield  {author} {\bibinfo {author} {\bibfnamefont {C.}~\bibnamefont {Lagoin}}, \bibinfo {author} {\bibfnamefont {U.}~\bibnamefont {Bhattacharya}}, \bibinfo {author} {\bibfnamefont {T.}~\bibnamefont {Grass}}, \bibinfo {author} {\bibfnamefont {R.~W.}\ \bibnamefont {Chhajlany}}, \bibinfo {author} {\bibfnamefont {T.}~\bibnamefont {Salamon}}, \bibinfo {author} {\bibfnamefont {K.}~\bibnamefont {Baldwin}}, \bibinfo {author} {\bibfnamefont {L.}~\bibnamefont {Pfeiffer}}, \bibinfo {author} {\bibfnamefont {M.}~\bibnamefont {Lewenstein}}, \bibinfo {author} {\bibfnamefont {M.}~\bibnamefont {Holzmann}},\ and\ \bibinfo {author} {\bibfnamefont {F.}~\bibnamefont {Dubin}},\ }\bibfield  {title} {\bibinfo {title} {Extended {{Bose}}--{{Hubbard}} model with dipolar excitons},\ }\href {https://doi.org/10.1038/s41586-022-05123-z} {\bibfield  {journal} {\bibinfo  {journal} {Nature}\ }\textbf {\bibinfo {volume} {609}},\ \bibinfo {pages} {485} (\bibinfo {year} {2022}{\natexlab{b}})}\BibitemShut {NoStop}%
\bibitem [{\citenamefont {Xiong}\ \emph {et~al.}(2023)\citenamefont {Xiong}, \citenamefont {Nie}, \citenamefont {Brantly}, \citenamefont {Hays}, \citenamefont {Sailus}, \citenamefont {Watanabe}, \citenamefont {Taniguchi}, \citenamefont {Tongay},\ and\ \citenamefont {Jin}}]{xiong_correlated_2023}%
  \BibitemOpen
  \bibfield  {author} {\bibinfo {author} {\bibfnamefont {R.}~\bibnamefont {Xiong}}, \bibinfo {author} {\bibfnamefont {J.~H.}\ \bibnamefont {Nie}}, \bibinfo {author} {\bibfnamefont {S.~L.}\ \bibnamefont {Brantly}}, \bibinfo {author} {\bibfnamefont {P.}~\bibnamefont {Hays}}, \bibinfo {author} {\bibfnamefont {R.}~\bibnamefont {Sailus}}, \bibinfo {author} {\bibfnamefont {K.}~\bibnamefont {Watanabe}}, \bibinfo {author} {\bibfnamefont {T.}~\bibnamefont {Taniguchi}}, \bibinfo {author} {\bibfnamefont {S.}~\bibnamefont {Tongay}},\ and\ \bibinfo {author} {\bibfnamefont {C.}~\bibnamefont {Jin}},\ }\bibfield  {title} {\bibinfo {title} {Correlated insulator of excitons in {{WSe2}}/{{WS2}} moir{\'e} superlattices},\ }\href {https://doi.org/10.1126/science.add5574} {\bibfield  {journal} {\bibinfo  {journal} {Science}\ }\textbf {\bibinfo {volume} {380}},\ \bibinfo {pages} {860} (\bibinfo {year} {2023})}\BibitemShut {NoStop}%
\bibitem [{\citenamefont {Ben~Mhenni}\ \emph {et~al.}(2024)\citenamefont {Ben~Mhenni}, \citenamefont {Kadow}, \citenamefont {Metelski}, \citenamefont {Paulus}, \citenamefont {Dijkstra}, \citenamefont {Watanabe}, \citenamefont {Taniguchi}, \citenamefont {Tongay}, \citenamefont {Barbone}, \citenamefont {Finley}, \citenamefont {Knap},\ and\ \citenamefont {Wilson}}]{ben_mhenni_gate-tunable_2024}%
  \BibitemOpen
  \bibfield  {author} {\bibinfo {author} {\bibfnamefont {A.}~\bibnamefont {Ben~Mhenni}}, \bibinfo {author} {\bibfnamefont {W.}~\bibnamefont {Kadow}}, \bibinfo {author} {\bibfnamefont {M.~J.}\ \bibnamefont {Metelski}}, \bibinfo {author} {\bibfnamefont {A.~O.}\ \bibnamefont {Paulus}}, \bibinfo {author} {\bibfnamefont {A.}~\bibnamefont {Dijkstra}}, \bibinfo {author} {\bibfnamefont {K.}~\bibnamefont {Watanabe}}, \bibinfo {author} {\bibfnamefont {T.}~\bibnamefont {Taniguchi}}, \bibinfo {author} {\bibfnamefont {S.~A.}\ \bibnamefont {Tongay}}, \bibinfo {author} {\bibfnamefont {M.}~\bibnamefont {Barbone}}, \bibinfo {author} {\bibfnamefont {J.~J.}\ \bibnamefont {Finley}}, \bibinfo {author} {\bibfnamefont {M.}~\bibnamefont {Knap}},\ and\ \bibinfo {author} {\bibfnamefont {N.~P.}\ \bibnamefont {Wilson}},\ }\href {https://doi.org/10.48550/arXiv.2410.07308} {\bibinfo {title} {Gate-tunable {{Bose-Fermi}} mixture in a strongly correlated moir{\'e} bilayer electron system}} (\bibinfo {year} {2024}),\ \Eprint
  {https://arxiv.org/abs/2410.07308} {arXiv:2410.07308 [cond-mat]} \BibitemShut {NoStop}%
\bibitem [{\citenamefont {Gao}\ \emph {et~al.}(2024)\citenamefont {Gao}, \citenamefont {{Su{\'a}rez-Forero}}, \citenamefont {Sarkar}, \citenamefont {Huang}, \citenamefont {Session}, \citenamefont {Mehrabad}, \citenamefont {Ni}, \citenamefont {Xie}, \citenamefont {Upadhyay}, \citenamefont {Vannucci}, \citenamefont {Mittal}, \citenamefont {Watanabe}, \citenamefont {Taniguchi}, \citenamefont {Imamoglu}, \citenamefont {Zhou},\ and\ \citenamefont {Hafezi}}]{gao_excitonic_2024}%
  \BibitemOpen
  \bibfield  {author} {\bibinfo {author} {\bibfnamefont {B.}~\bibnamefont {Gao}}, \bibinfo {author} {\bibfnamefont {D.~G.}\ \bibnamefont {{Su{\'a}rez-Forero}}}, \bibinfo {author} {\bibfnamefont {S.}~\bibnamefont {Sarkar}}, \bibinfo {author} {\bibfnamefont {T.-S.}\ \bibnamefont {Huang}}, \bibinfo {author} {\bibfnamefont {D.}~\bibnamefont {Session}}, \bibinfo {author} {\bibfnamefont {M.~J.}\ \bibnamefont {Mehrabad}}, \bibinfo {author} {\bibfnamefont {R.}~\bibnamefont {Ni}}, \bibinfo {author} {\bibfnamefont {M.}~\bibnamefont {Xie}}, \bibinfo {author} {\bibfnamefont {P.}~\bibnamefont {Upadhyay}}, \bibinfo {author} {\bibfnamefont {J.}~\bibnamefont {Vannucci}}, \bibinfo {author} {\bibfnamefont {S.}~\bibnamefont {Mittal}}, \bibinfo {author} {\bibfnamefont {K.}~\bibnamefont {Watanabe}}, \bibinfo {author} {\bibfnamefont {T.}~\bibnamefont {Taniguchi}}, \bibinfo {author} {\bibfnamefont {A.}~\bibnamefont {Imamoglu}}, \bibinfo {author} {\bibfnamefont {Y.}~\bibnamefont {Zhou}},\ and\ \bibinfo {author} {\bibfnamefont
  {M.}~\bibnamefont {Hafezi}},\ }\bibfield  {title} {\bibinfo {title} {Excitonic {{Mott}} insulator in a {{Bose-Fermi-Hubbard}} system of moir{\'e} {{WS2}}/{{WSe2}} heterobilayer},\ }\href {https://doi.org/10.1038/s41467-024-46616-x} {\bibfield  {journal} {\bibinfo  {journal} {Nat Commun}\ }\textbf {\bibinfo {volume} {15}},\ \bibinfo {pages} {2305} (\bibinfo {year} {2024})}\BibitemShut {NoStop}%
\bibitem [{\citenamefont {Zhang}\ \emph {et~al.}(2021)\citenamefont {Zhang}, \citenamefont {Sheng},\ and\ \citenamefont {Vishwanath}}]{zhang_su4_2021}%
  \BibitemOpen
  \bibfield  {author} {\bibinfo {author} {\bibfnamefont {Y.-H.}\ \bibnamefont {Zhang}}, \bibinfo {author} {\bibfnamefont {D.~N.}\ \bibnamefont {Sheng}},\ and\ \bibinfo {author} {\bibfnamefont {A.}~\bibnamefont {Vishwanath}},\ }\bibfield  {title} {\bibinfo {title} {{{SU}}(4) {{Chiral Spin Liquid}}, {{Exciton Supersolid}}, and {{Electric Detection}} in {{Moir}}{\textbackslash}'e {{Bilayers}}},\ }\href {https://doi.org/10.1103/PhysRevLett.127.247701} {\bibfield  {journal} {\bibinfo  {journal} {Phys. Rev. Lett.}\ }\textbf {\bibinfo {volume} {127}},\ \bibinfo {pages} {247701} (\bibinfo {year} {2021})}\BibitemShut {NoStop}%
\bibitem [{\citenamefont {Zeng}\ \emph {et~al.}(2023)\citenamefont {Zeng}, \citenamefont {Xia}, \citenamefont {Dery}, \citenamefont {Watanabe}, \citenamefont {Taniguchi}, \citenamefont {Shan},\ and\ \citenamefont {Mak}}]{zeng_exciton_2023}%
  \BibitemOpen
  \bibfield  {author} {\bibinfo {author} {\bibfnamefont {Y.}~\bibnamefont {Zeng}}, \bibinfo {author} {\bibfnamefont {Z.}~\bibnamefont {Xia}}, \bibinfo {author} {\bibfnamefont {R.}~\bibnamefont {Dery}}, \bibinfo {author} {\bibfnamefont {K.}~\bibnamefont {Watanabe}}, \bibinfo {author} {\bibfnamefont {T.}~\bibnamefont {Taniguchi}}, \bibinfo {author} {\bibfnamefont {J.}~\bibnamefont {Shan}},\ and\ \bibinfo {author} {\bibfnamefont {K.~F.}\ \bibnamefont {Mak}},\ }\bibfield  {title} {\bibinfo {title} {Exciton density waves in {{Coulomb-coupled}} dual moir{\'e} lattices},\ }\href {https://doi.org/10.1038/s41563-022-01454-4} {\bibfield  {journal} {\bibinfo  {journal} {Nat. Mater.}\ }\textbf {\bibinfo {volume} {22}},\ \bibinfo {pages} {175} (\bibinfo {year} {2023})}\BibitemShut {NoStop}%
\bibitem [{\citenamefont {Eisenstein}\ and\ \citenamefont {MacDonald}(2004)}]{eisenstein_boseeinstein_2004}%
  \BibitemOpen
  \bibfield  {author} {\bibinfo {author} {\bibfnamefont {J.~P.}\ \bibnamefont {Eisenstein}}\ and\ \bibinfo {author} {\bibfnamefont {A.~H.}\ \bibnamefont {MacDonald}},\ }\bibfield  {title} {\bibinfo {title} {Bose--{{Einstein}} condensation of excitons in bilayer electron systems},\ }\href {https://doi.org/10.1038/nature03081} {\bibfield  {journal} {\bibinfo  {journal} {Nature}\ }\textbf {\bibinfo {volume} {432}},\ \bibinfo {pages} {691} (\bibinfo {year} {2004})}\BibitemShut {NoStop}%
\bibitem [{\citenamefont {Fogler}\ \emph {et~al.}(2014)\citenamefont {Fogler}, \citenamefont {Butov},\ and\ \citenamefont {Novoselov}}]{fogler_high-temperature_2014}%
  \BibitemOpen
  \bibfield  {author} {\bibinfo {author} {\bibfnamefont {M.~M.}\ \bibnamefont {Fogler}}, \bibinfo {author} {\bibfnamefont {L.~V.}\ \bibnamefont {Butov}},\ and\ \bibinfo {author} {\bibfnamefont {K.~S.}\ \bibnamefont {Novoselov}},\ }\bibfield  {title} {\bibinfo {title} {High-temperature superfluidity with indirect excitons in van der {{Waals}} heterostructures},\ }\href {https://doi.org/10.1038/ncomms5555} {\bibfield  {journal} {\bibinfo  {journal} {Nat Commun}\ }\textbf {\bibinfo {volume} {5}},\ \bibinfo {pages} {4555} (\bibinfo {year} {2014})}\BibitemShut {NoStop}%
\bibitem [{\citenamefont {Ma}\ \emph {et~al.}(2021)\citenamefont {Ma}, \citenamefont {Nguyen}, \citenamefont {Wang}, \citenamefont {Zeng}, \citenamefont {Watanabe}, \citenamefont {Taniguchi}, \citenamefont {MacDonald}, \citenamefont {Mak},\ and\ \citenamefont {Shan}}]{ma_strongly_2021}%
  \BibitemOpen
  \bibfield  {author} {\bibinfo {author} {\bibfnamefont {L.}~\bibnamefont {Ma}}, \bibinfo {author} {\bibfnamefont {P.~X.}\ \bibnamefont {Nguyen}}, \bibinfo {author} {\bibfnamefont {Z.}~\bibnamefont {Wang}}, \bibinfo {author} {\bibfnamefont {Y.}~\bibnamefont {Zeng}}, \bibinfo {author} {\bibfnamefont {K.}~\bibnamefont {Watanabe}}, \bibinfo {author} {\bibfnamefont {T.}~\bibnamefont {Taniguchi}}, \bibinfo {author} {\bibfnamefont {A.~H.}\ \bibnamefont {MacDonald}}, \bibinfo {author} {\bibfnamefont {K.~F.}\ \bibnamefont {Mak}},\ and\ \bibinfo {author} {\bibfnamefont {J.}~\bibnamefont {Shan}},\ }\bibfield  {title} {\bibinfo {title} {Strongly correlated excitonic insulator in atomic double layers},\ }\href {https://doi.org/10.1038/s41586-021-03947-9} {\bibfield  {journal} {\bibinfo  {journal} {Nature}\ }\textbf {\bibinfo {volume} {598}},\ \bibinfo {pages} {585} (\bibinfo {year} {2021})}\BibitemShut {NoStop}%
\bibitem [{\citenamefont {Jauregui}\ \emph {et~al.}(2019)\citenamefont {Jauregui}, \citenamefont {Joe}, \citenamefont {Pistunova}, \citenamefont {Wild}, \citenamefont {High}, \citenamefont {Zhou}, \citenamefont {Scuri}, \citenamefont {De~Greve}, \citenamefont {Sushko}, \citenamefont {Yu}, \citenamefont {Taniguchi}, \citenamefont {Watanabe}, \citenamefont {Needleman}, \citenamefont {Lukin}, \citenamefont {Park},\ and\ \citenamefont {Kim}}]{jauregui_electrical_2019}%
  \BibitemOpen
  \bibfield  {author} {\bibinfo {author} {\bibfnamefont {L.~A.}\ \bibnamefont {Jauregui}}, \bibinfo {author} {\bibfnamefont {A.~Y.}\ \bibnamefont {Joe}}, \bibinfo {author} {\bibfnamefont {K.}~\bibnamefont {Pistunova}}, \bibinfo {author} {\bibfnamefont {D.~S.}\ \bibnamefont {Wild}}, \bibinfo {author} {\bibfnamefont {A.~A.}\ \bibnamefont {High}}, \bibinfo {author} {\bibfnamefont {Y.}~\bibnamefont {Zhou}}, \bibinfo {author} {\bibfnamefont {G.}~\bibnamefont {Scuri}}, \bibinfo {author} {\bibfnamefont {K.}~\bibnamefont {De~Greve}}, \bibinfo {author} {\bibfnamefont {A.}~\bibnamefont {Sushko}}, \bibinfo {author} {\bibfnamefont {C.-H.}\ \bibnamefont {Yu}}, \bibinfo {author} {\bibfnamefont {T.}~\bibnamefont {Taniguchi}}, \bibinfo {author} {\bibfnamefont {K.}~\bibnamefont {Watanabe}}, \bibinfo {author} {\bibfnamefont {D.~J.}\ \bibnamefont {Needleman}}, \bibinfo {author} {\bibfnamefont {M.~D.}\ \bibnamefont {Lukin}}, \bibinfo {author} {\bibfnamefont {H.}~\bibnamefont {Park}},\ and\ \bibinfo {author} {\bibfnamefont
  {P.}~\bibnamefont {Kim}},\ }\bibfield  {title} {\bibinfo {title} {Electrical control of interlayer exciton dynamics in atomically thin heterostructures},\ }\href {https://doi.org/10.1126/science.aaw4194} {\bibfield  {journal} {\bibinfo  {journal} {Science}\ }\textbf {\bibinfo {volume} {366}},\ \bibinfo {pages} {870} (\bibinfo {year} {2019})}\BibitemShut {NoStop}%
\bibitem [{\citenamefont {Wang}\ \emph {et~al.}(2019)\citenamefont {Wang}, \citenamefont {Rhodes}, \citenamefont {Watanabe}, \citenamefont {Taniguchi}, \citenamefont {Hone}, \citenamefont {Shan},\ and\ \citenamefont {Mak}}]{wang_evidence_2019}%
  \BibitemOpen
  \bibfield  {author} {\bibinfo {author} {\bibfnamefont {Z.}~\bibnamefont {Wang}}, \bibinfo {author} {\bibfnamefont {D.~A.}\ \bibnamefont {Rhodes}}, \bibinfo {author} {\bibfnamefont {K.}~\bibnamefont {Watanabe}}, \bibinfo {author} {\bibfnamefont {T.}~\bibnamefont {Taniguchi}}, \bibinfo {author} {\bibfnamefont {J.~C.}\ \bibnamefont {Hone}}, \bibinfo {author} {\bibfnamefont {J.}~\bibnamefont {Shan}},\ and\ \bibinfo {author} {\bibfnamefont {K.~F.}\ \bibnamefont {Mak}},\ }\bibfield  {title} {\bibinfo {title} {Evidence of high-temperature exciton condensation in two-dimensional atomic double layers},\ }\href {https://doi.org/10.1038/s41586-019-1591-7} {\bibfield  {journal} {\bibinfo  {journal} {Nature}\ }\textbf {\bibinfo {volume} {574}},\ \bibinfo {pages} {76} (\bibinfo {year} {2019})}\BibitemShut {NoStop}%
\bibitem [{\citenamefont {Liu}\ \emph {et~al.}(2022)\citenamefont {Liu}, \citenamefont {Li}, \citenamefont {Watanabe}, \citenamefont {Taniguchi}, \citenamefont {Hone}, \citenamefont {Halperin}, \citenamefont {Kim},\ and\ \citenamefont {Dean}}]{liu_crossover_2022}%
  \BibitemOpen
  \bibfield  {author} {\bibinfo {author} {\bibfnamefont {X.}~\bibnamefont {Liu}}, \bibinfo {author} {\bibfnamefont {J.~I.~A.}\ \bibnamefont {Li}}, \bibinfo {author} {\bibfnamefont {K.}~\bibnamefont {Watanabe}}, \bibinfo {author} {\bibfnamefont {T.}~\bibnamefont {Taniguchi}}, \bibinfo {author} {\bibfnamefont {J.}~\bibnamefont {Hone}}, \bibinfo {author} {\bibfnamefont {B.~I.}\ \bibnamefont {Halperin}}, \bibinfo {author} {\bibfnamefont {P.}~\bibnamefont {Kim}},\ and\ \bibinfo {author} {\bibfnamefont {C.~R.}\ \bibnamefont {Dean}},\ }\bibfield  {title} {\bibinfo {title} {Crossover between strongly coupled and weakly coupled exciton superfluids},\ }\href {https://doi.org/10.1126/science.abg1110} {\bibfield  {journal} {\bibinfo  {journal} {Science}\ }\textbf {\bibinfo {volume} {375}},\ \bibinfo {pages} {205} (\bibinfo {year} {2022})}\BibitemShut {NoStop}%
\bibitem [{\citenamefont {Tang}\ \emph {et~al.}(2020)\citenamefont {Tang}, \citenamefont {Li}, \citenamefont {Li}, \citenamefont {Xu}, \citenamefont {Liu}, \citenamefont {Barmak}, \citenamefont {Watanabe}, \citenamefont {Taniguchi}, \citenamefont {MacDonald}, \citenamefont {Shan},\ and\ \citenamefont {Mak}}]{tang_simulation_2020}%
  \BibitemOpen
  \bibfield  {author} {\bibinfo {author} {\bibfnamefont {Y.}~\bibnamefont {Tang}}, \bibinfo {author} {\bibfnamefont {L.}~\bibnamefont {Li}}, \bibinfo {author} {\bibfnamefont {T.}~\bibnamefont {Li}}, \bibinfo {author} {\bibfnamefont {Y.}~\bibnamefont {Xu}}, \bibinfo {author} {\bibfnamefont {S.}~\bibnamefont {Liu}}, \bibinfo {author} {\bibfnamefont {K.}~\bibnamefont {Barmak}}, \bibinfo {author} {\bibfnamefont {K.}~\bibnamefont {Watanabe}}, \bibinfo {author} {\bibfnamefont {T.}~\bibnamefont {Taniguchi}}, \bibinfo {author} {\bibfnamefont {A.~H.}\ \bibnamefont {MacDonald}}, \bibinfo {author} {\bibfnamefont {J.}~\bibnamefont {Shan}},\ and\ \bibinfo {author} {\bibfnamefont {K.~F.}\ \bibnamefont {Mak}},\ }\bibfield  {title} {\bibinfo {title} {Simulation of {{Hubbard}} model physics in {{WSe2}}/{{WS2}} moir{\'e} superlattices},\ }\href {https://doi.org/10.1038/s41586-020-2085-3} {\bibfield  {journal} {\bibinfo  {journal} {Nature}\ }\textbf {\bibinfo {volume} {579}},\ \bibinfo {pages} {353} (\bibinfo {year}
  {2020})}\BibitemShut {NoStop}%
\bibitem [{\citenamefont {Regan}\ \emph {et~al.}(2020)\citenamefont {Regan}, \citenamefont {Wang}, \citenamefont {Jin}, \citenamefont {Bakti~Utama}, \citenamefont {Gao}, \citenamefont {Wei}, \citenamefont {Zhao}, \citenamefont {Zhao}, \citenamefont {Zhang}, \citenamefont {Yumigeta}, \citenamefont {Blei}, \citenamefont {Carlstr{\"o}m}, \citenamefont {Watanabe}, \citenamefont {Taniguchi}, \citenamefont {Tongay}, \citenamefont {Crommie}, \citenamefont {Zettl},\ and\ \citenamefont {Wang}}]{regan_mott_2020}%
  \BibitemOpen
  \bibfield  {author} {\bibinfo {author} {\bibfnamefont {E.~C.}\ \bibnamefont {Regan}}, \bibinfo {author} {\bibfnamefont {D.}~\bibnamefont {Wang}}, \bibinfo {author} {\bibfnamefont {C.}~\bibnamefont {Jin}}, \bibinfo {author} {\bibfnamefont {M.~I.}\ \bibnamefont {Bakti~Utama}}, \bibinfo {author} {\bibfnamefont {B.}~\bibnamefont {Gao}}, \bibinfo {author} {\bibfnamefont {X.}~\bibnamefont {Wei}}, \bibinfo {author} {\bibfnamefont {S.}~\bibnamefont {Zhao}}, \bibinfo {author} {\bibfnamefont {W.}~\bibnamefont {Zhao}}, \bibinfo {author} {\bibfnamefont {Z.}~\bibnamefont {Zhang}}, \bibinfo {author} {\bibfnamefont {K.}~\bibnamefont {Yumigeta}}, \bibinfo {author} {\bibfnamefont {M.}~\bibnamefont {Blei}}, \bibinfo {author} {\bibfnamefont {J.~D.}\ \bibnamefont {Carlstr{\"o}m}}, \bibinfo {author} {\bibfnamefont {K.}~\bibnamefont {Watanabe}}, \bibinfo {author} {\bibfnamefont {T.}~\bibnamefont {Taniguchi}}, \bibinfo {author} {\bibfnamefont {S.}~\bibnamefont {Tongay}}, \bibinfo {author} {\bibfnamefont {M.}~\bibnamefont
  {Crommie}}, \bibinfo {author} {\bibfnamefont {A.}~\bibnamefont {Zettl}},\ and\ \bibinfo {author} {\bibfnamefont {F.}~\bibnamefont {Wang}},\ }\bibfield  {title} {\bibinfo {title} {Mott and generalized {{Wigner}} crystal states in {{WSe2}}/{{WS2}} moir{\'e} superlattices},\ }\href {https://doi.org/10.1038/s41586-020-2092-4} {\bibfield  {journal} {\bibinfo  {journal} {Nature}\ }\textbf {\bibinfo {volume} {579}},\ \bibinfo {pages} {359} (\bibinfo {year} {2020})}\BibitemShut {NoStop}%
\bibitem [{\citenamefont {Xu}\ \emph {et~al.}(2020)\citenamefont {Xu}, \citenamefont {Liu}, \citenamefont {Rhodes}, \citenamefont {Watanabe}, \citenamefont {Taniguchi}, \citenamefont {Hone}, \citenamefont {Elser}, \citenamefont {Mak},\ and\ \citenamefont {Shan}}]{xu_correlated_2020}%
  \BibitemOpen
  \bibfield  {author} {\bibinfo {author} {\bibfnamefont {Y.}~\bibnamefont {Xu}}, \bibinfo {author} {\bibfnamefont {S.}~\bibnamefont {Liu}}, \bibinfo {author} {\bibfnamefont {D.~A.}\ \bibnamefont {Rhodes}}, \bibinfo {author} {\bibfnamefont {K.}~\bibnamefont {Watanabe}}, \bibinfo {author} {\bibfnamefont {T.}~\bibnamefont {Taniguchi}}, \bibinfo {author} {\bibfnamefont {J.}~\bibnamefont {Hone}}, \bibinfo {author} {\bibfnamefont {V.}~\bibnamefont {Elser}}, \bibinfo {author} {\bibfnamefont {K.~F.}\ \bibnamefont {Mak}},\ and\ \bibinfo {author} {\bibfnamefont {J.}~\bibnamefont {Shan}},\ }\bibfield  {title} {\bibinfo {title} {Correlated insulating states at fractional fillings of moir{\'e} superlattices},\ }\href {https://doi.org/10.1038/s41586-020-2868-6} {\bibfield  {journal} {\bibinfo  {journal} {Nature}\ }\textbf {\bibinfo {volume} {587}},\ \bibinfo {pages} {214} (\bibinfo {year} {2020})}\BibitemShut {NoStop}%
\bibitem [{\citenamefont {Wang}\ \emph {et~al.}(2020)\citenamefont {Wang}, \citenamefont {Shih}, \citenamefont {Ghiotto}, \citenamefont {Xian}, \citenamefont {Rhodes}, \citenamefont {Tan}, \citenamefont {Claassen}, \citenamefont {Kennes}, \citenamefont {Bai}, \citenamefont {Kim}, \citenamefont {Watanabe}, \citenamefont {Taniguchi}, \citenamefont {Zhu}, \citenamefont {Hone}, \citenamefont {Rubio}, \citenamefont {Pasupathy},\ and\ \citenamefont {Dean}}]{wang_correlated_2020}%
  \BibitemOpen
  \bibfield  {author} {\bibinfo {author} {\bibfnamefont {L.}~\bibnamefont {Wang}}, \bibinfo {author} {\bibfnamefont {E.-M.}\ \bibnamefont {Shih}}, \bibinfo {author} {\bibfnamefont {A.}~\bibnamefont {Ghiotto}}, \bibinfo {author} {\bibfnamefont {L.}~\bibnamefont {Xian}}, \bibinfo {author} {\bibfnamefont {D.~A.}\ \bibnamefont {Rhodes}}, \bibinfo {author} {\bibfnamefont {C.}~\bibnamefont {Tan}}, \bibinfo {author} {\bibfnamefont {M.}~\bibnamefont {Claassen}}, \bibinfo {author} {\bibfnamefont {D.~M.}\ \bibnamefont {Kennes}}, \bibinfo {author} {\bibfnamefont {Y.}~\bibnamefont {Bai}}, \bibinfo {author} {\bibfnamefont {B.}~\bibnamefont {Kim}}, \bibinfo {author} {\bibfnamefont {K.}~\bibnamefont {Watanabe}}, \bibinfo {author} {\bibfnamefont {T.}~\bibnamefont {Taniguchi}}, \bibinfo {author} {\bibfnamefont {X.}~\bibnamefont {Zhu}}, \bibinfo {author} {\bibfnamefont {J.}~\bibnamefont {Hone}}, \bibinfo {author} {\bibfnamefont {A.}~\bibnamefont {Rubio}}, \bibinfo {author} {\bibfnamefont {A.~N.}\ \bibnamefont {Pasupathy}},\
  and\ \bibinfo {author} {\bibfnamefont {C.~R.}\ \bibnamefont {Dean}},\ }\bibfield  {title} {\bibinfo {title} {Correlated electronic phases in twisted bilayer transition metal dichalcogenides},\ }\href {https://doi.org/10.1038/s41563-020-0708-6} {\bibfield  {journal} {\bibinfo  {journal} {Nat. Mater.}\ }\textbf {\bibinfo {volume} {19}},\ \bibinfo {pages} {861} (\bibinfo {year} {2020})}\BibitemShut {NoStop}%
\bibitem [{\citenamefont {Shimazaki}\ \emph {et~al.}(2020)\citenamefont {Shimazaki}, \citenamefont {Schwartz}, \citenamefont {Watanabe}, \citenamefont {Taniguchi}, \citenamefont {Kroner},\ and\ \citenamefont {Imamo{\u g}lu}}]{shimazaki_strongly_2020}%
  \BibitemOpen
  \bibfield  {author} {\bibinfo {author} {\bibfnamefont {Y.}~\bibnamefont {Shimazaki}}, \bibinfo {author} {\bibfnamefont {I.}~\bibnamefont {Schwartz}}, \bibinfo {author} {\bibfnamefont {K.}~\bibnamefont {Watanabe}}, \bibinfo {author} {\bibfnamefont {T.}~\bibnamefont {Taniguchi}}, \bibinfo {author} {\bibfnamefont {M.}~\bibnamefont {Kroner}},\ and\ \bibinfo {author} {\bibfnamefont {A.}~\bibnamefont {Imamo{\u g}lu}},\ }\bibfield  {title} {\bibinfo {title} {Strongly correlated electrons and hybrid excitons in a moir{\'e} heterostructure},\ }\href {https://doi.org/10.1038/s41586-020-2191-2} {\bibfield  {journal} {\bibinfo  {journal} {Nature}\ }\textbf {\bibinfo {volume} {580}},\ \bibinfo {pages} {472} (\bibinfo {year} {2020})}\BibitemShut {NoStop}%
\bibitem [{\citenamefont {Wilson}\ \emph {et~al.}(2021)\citenamefont {Wilson}, \citenamefont {Yao}, \citenamefont {Shan},\ and\ \citenamefont {Xu}}]{wilson_excitons_2021}%
  \BibitemOpen
  \bibfield  {author} {\bibinfo {author} {\bibfnamefont {N.~P.}\ \bibnamefont {Wilson}}, \bibinfo {author} {\bibfnamefont {W.}~\bibnamefont {Yao}}, \bibinfo {author} {\bibfnamefont {J.}~\bibnamefont {Shan}},\ and\ \bibinfo {author} {\bibfnamefont {X.}~\bibnamefont {Xu}},\ }\bibfield  {title} {\bibinfo {title} {Excitons and emergent quantum phenomena in stacked {{2D}} semiconductors},\ }\href {https://doi.org/10.1038/s41586-021-03979-1} {\bibfield  {journal} {\bibinfo  {journal} {Nature}\ }\textbf {\bibinfo {volume} {599}},\ \bibinfo {pages} {383} (\bibinfo {year} {2021})}\BibitemShut {NoStop}%
\bibitem [{\citenamefont {Zeng}\ \emph {et~al.}(2022)\citenamefont {Zeng}, \citenamefont {Wei},\ and\ \citenamefont {MacDonald}}]{zeng_layer_2022}%
  \BibitemOpen
  \bibfield  {author} {\bibinfo {author} {\bibfnamefont {Y.}~\bibnamefont {Zeng}}, \bibinfo {author} {\bibfnamefont {N.}~\bibnamefont {Wei}},\ and\ \bibinfo {author} {\bibfnamefont {A.~H.}\ \bibnamefont {MacDonald}},\ }\bibfield  {title} {\bibinfo {title} {Layer pseudospin magnetism in a transition metal dichalcogenide double-moir{\textbackslash}'e system},\ }\href {https://doi.org/10.1103/PhysRevB.106.165105} {\bibfield  {journal} {\bibinfo  {journal} {Phys. Rev. B}\ }\textbf {\bibinfo {volume} {106}},\ \bibinfo {pages} {165105} (\bibinfo {year} {2022})}\BibitemShut {NoStop}%
\bibitem [{\citenamefont {Zhang}(2022)}]{zhang_doping_2022}%
  \BibitemOpen
  \bibfield  {author} {\bibinfo {author} {\bibfnamefont {Y.-H.}\ \bibnamefont {Zhang}},\ }\bibfield  {title} {\bibinfo {title} {Doping a {{Mott}} insulator with excitons in a moir{\textbackslash}'e bilayer: {{Fractional}} superfluid, neutral {{Fermi}} surface, and {{Mott}} transition},\ }\href {https://doi.org/10.1103/PhysRevB.106.195120} {\bibfield  {journal} {\bibinfo  {journal} {Phys. Rev. B}\ }\textbf {\bibinfo {volume} {106}},\ \bibinfo {pages} {195120} (\bibinfo {year} {2022})}\BibitemShut {NoStop}%
\bibitem [{\citenamefont {Xie}\ \emph {et~al.}(2024)\citenamefont {Xie}, \citenamefont {Hafezi},\ and\ \citenamefont {Das~Sarma}}]{xie_long-lived_2024}%
  \BibitemOpen
  \bibfield  {author} {\bibinfo {author} {\bibfnamefont {M.}~\bibnamefont {Xie}}, \bibinfo {author} {\bibfnamefont {M.}~\bibnamefont {Hafezi}},\ and\ \bibinfo {author} {\bibfnamefont {S.}~\bibnamefont {Das~Sarma}},\ }\bibfield  {title} {\bibinfo {title} {Long-{{Lived Topological Flatband Excitons}} in {{Semiconductor Moir}}{\textbackslash}'e {{Heterostructures}}: {{A Bosonic Kane-Mele Model Platform}}},\ }\href {https://doi.org/10.1103/PhysRevLett.133.136403} {\bibfield  {journal} {\bibinfo  {journal} {Phys. Rev. Lett.}\ }\textbf {\bibinfo {volume} {133}},\ \bibinfo {pages} {136403} (\bibinfo {year} {2024})}\BibitemShut {NoStop}%
\bibitem [{\citenamefont {Zhang}\ \emph {et~al.}(2025)\citenamefont {Zhang}, \citenamefont {Yao},\ and\ \citenamefont {Yu}}]{zhang_engineering_2025}%
  \BibitemOpen
  \bibfield  {author} {\bibinfo {author} {\bibfnamefont {N.}~\bibnamefont {Zhang}}, \bibinfo {author} {\bibfnamefont {W.}~\bibnamefont {Yao}},\ and\ \bibinfo {author} {\bibfnamefont {H.}~\bibnamefont {Yu}},\ }\href {https://doi.org/10.48550/arXiv.2505.04597} {\bibinfo {title} {Engineering topological exciton structures in two-dimensional semiconductors by a periodic electrostatic potential}} (\bibinfo {year} {2025}),\ \Eprint {https://arxiv.org/abs/2505.04597} {arXiv:2505.04597 [cond-mat]} \BibitemShut {NoStop}%
\bibitem [{\citenamefont {Zhao}\ \emph {et~al.}(2021)\citenamefont {Zhao}, \citenamefont {Xiao},\ and\ \citenamefont {Yao}}]{zhao_universal_2021}%
  \BibitemOpen
  \bibfield  {author} {\bibinfo {author} {\bibfnamefont {P.}~\bibnamefont {Zhao}}, \bibinfo {author} {\bibfnamefont {C.}~\bibnamefont {Xiao}},\ and\ \bibinfo {author} {\bibfnamefont {W.}~\bibnamefont {Yao}},\ }\bibfield  {title} {\bibinfo {title} {Universal superlattice potential for {{2D}} materials from twisted interface inside h-{{BN}} substrate},\ }\href {https://doi.org/10.1038/s41699-021-00221-4} {\bibfield  {journal} {\bibinfo  {journal} {npj 2D Mater Appl}\ }\textbf {\bibinfo {volume} {5}},\ \bibinfo {pages} {1} (\bibinfo {year} {2021})}\BibitemShut {NoStop}%
\bibitem [{\citenamefont {Kim}\ \emph {et~al.}(2024)\citenamefont {Kim}, \citenamefont {Dominguez}, \citenamefont {{Mayorga-Luna}}, \citenamefont {Ye}, \citenamefont {Embley}, \citenamefont {Tan}, \citenamefont {Ni}, \citenamefont {Liu}, \citenamefont {Ford}, \citenamefont {Gao}, \citenamefont {Arash}, \citenamefont {Watanabe}, \citenamefont {Taniguchi}, \citenamefont {Kim}, \citenamefont {Shih}, \citenamefont {Lai}, \citenamefont {Yao}, \citenamefont {Yang}, \citenamefont {Li},\ and\ \citenamefont {Miyahara}}]{kim_electrostatic_2024}%
  \BibitemOpen
  \bibfield  {author} {\bibinfo {author} {\bibfnamefont {D.~S.}\ \bibnamefont {Kim}}, \bibinfo {author} {\bibfnamefont {R.~C.}\ \bibnamefont {Dominguez}}, \bibinfo {author} {\bibfnamefont {R.}~\bibnamefont {{Mayorga-Luna}}}, \bibinfo {author} {\bibfnamefont {D.}~\bibnamefont {Ye}}, \bibinfo {author} {\bibfnamefont {J.}~\bibnamefont {Embley}}, \bibinfo {author} {\bibfnamefont {T.}~\bibnamefont {Tan}}, \bibinfo {author} {\bibfnamefont {Y.}~\bibnamefont {Ni}}, \bibinfo {author} {\bibfnamefont {Z.}~\bibnamefont {Liu}}, \bibinfo {author} {\bibfnamefont {M.}~\bibnamefont {Ford}}, \bibinfo {author} {\bibfnamefont {F.~Y.}\ \bibnamefont {Gao}}, \bibinfo {author} {\bibfnamefont {S.}~\bibnamefont {Arash}}, \bibinfo {author} {\bibfnamefont {K.}~\bibnamefont {Watanabe}}, \bibinfo {author} {\bibfnamefont {T.}~\bibnamefont {Taniguchi}}, \bibinfo {author} {\bibfnamefont {S.}~\bibnamefont {Kim}}, \bibinfo {author} {\bibfnamefont {C.-K.}\ \bibnamefont {Shih}}, \bibinfo {author} {\bibfnamefont {K.}~\bibnamefont {Lai}}, \bibinfo
  {author} {\bibfnamefont {W.}~\bibnamefont {Yao}}, \bibinfo {author} {\bibfnamefont {L.}~\bibnamefont {Yang}}, \bibinfo {author} {\bibfnamefont {X.}~\bibnamefont {Li}},\ and\ \bibinfo {author} {\bibfnamefont {Y.}~\bibnamefont {Miyahara}},\ }\bibfield  {title} {\bibinfo {title} {Electrostatic moir{\'e} potential from twisted hexagonal boron nitride layers},\ }\href {https://doi.org/10.1038/s41563-023-01637-7} {\bibfield  {journal} {\bibinfo  {journal} {Nat. Mater.}\ }\textbf {\bibinfo {volume} {23}},\ \bibinfo {pages} {65} (\bibinfo {year} {2024})}\BibitemShut {NoStop}%
\bibitem [{\citenamefont {Wang}\ \emph {et~al.}(2025)\citenamefont {Wang}, \citenamefont {Xu}, \citenamefont {Aronson}, \citenamefont {Bennett}, \citenamefont {Paul}, \citenamefont {Crowley}, \citenamefont {Collignon}, \citenamefont {Watanabe}, \citenamefont {Taniguchi}, \citenamefont {Ashoori}, \citenamefont {Kaxiras}, \citenamefont {Zhang}, \citenamefont {{Jarillo-Herrero}},\ and\ \citenamefont {Yasuda}}]{wang_moire_2025}%
  \BibitemOpen
  \bibfield  {author} {\bibinfo {author} {\bibfnamefont {X.}~\bibnamefont {Wang}}, \bibinfo {author} {\bibfnamefont {C.}~\bibnamefont {Xu}}, \bibinfo {author} {\bibfnamefont {S.}~\bibnamefont {Aronson}}, \bibinfo {author} {\bibfnamefont {D.}~\bibnamefont {Bennett}}, \bibinfo {author} {\bibfnamefont {N.}~\bibnamefont {Paul}}, \bibinfo {author} {\bibfnamefont {P.~J.~D.}\ \bibnamefont {Crowley}}, \bibinfo {author} {\bibfnamefont {C.}~\bibnamefont {Collignon}}, \bibinfo {author} {\bibfnamefont {K.}~\bibnamefont {Watanabe}}, \bibinfo {author} {\bibfnamefont {T.}~\bibnamefont {Taniguchi}}, \bibinfo {author} {\bibfnamefont {R.}~\bibnamefont {Ashoori}}, \bibinfo {author} {\bibfnamefont {E.}~\bibnamefont {Kaxiras}}, \bibinfo {author} {\bibfnamefont {Y.}~\bibnamefont {Zhang}}, \bibinfo {author} {\bibfnamefont {P.}~\bibnamefont {{Jarillo-Herrero}}},\ and\ \bibinfo {author} {\bibfnamefont {K.}~\bibnamefont {Yasuda}},\ }\bibfield  {title} {\bibinfo {title} {Moir{\'e} band structure engineering using a twisted boron
  nitride substrate},\ }\href {https://doi.org/10.1038/s41467-024-55432-2} {\bibfield  {journal} {\bibinfo  {journal} {Nat Commun}\ }\textbf {\bibinfo {volume} {16}},\ \bibinfo {pages} {178} (\bibinfo {year} {2025})}\BibitemShut {NoStop}%
\bibitem [{\citenamefont {Li}\ and\ \citenamefont {Wu}(2017)}]{li_binary_2017}%
  \BibitemOpen
  \bibfield  {author} {\bibinfo {author} {\bibfnamefont {L.}~\bibnamefont {Li}}\ and\ \bibinfo {author} {\bibfnamefont {M.}~\bibnamefont {Wu}},\ }\bibfield  {title} {\bibinfo {title} {Binary {{Compound Bilayer}} and {{Multilayer}} with {{Vertical Polarizations}}: {{Two-Dimensional Ferroelectrics}}, {{Multiferroics}}, and {{Nanogenerators}}},\ }\href {https://doi.org/10.1021/acsnano.7b02756} {\bibfield  {journal} {\bibinfo  {journal} {ACS Nano}\ }\textbf {\bibinfo {volume} {11}},\ \bibinfo {pages} {6382} (\bibinfo {year} {2017})}\BibitemShut {NoStop}%
\bibitem [{\citenamefont {Constantinescu}\ \emph {et~al.}(2013)\citenamefont {Constantinescu}, \citenamefont {Kuc},\ and\ \citenamefont {Heine}}]{constantinescu_stacking_2013}%
  \BibitemOpen
  \bibfield  {author} {\bibinfo {author} {\bibfnamefont {G.}~\bibnamefont {Constantinescu}}, \bibinfo {author} {\bibfnamefont {A.}~\bibnamefont {Kuc}},\ and\ \bibinfo {author} {\bibfnamefont {T.}~\bibnamefont {Heine}},\ }\bibfield  {title} {\bibinfo {title} {Stacking in {{Bulk}} and {{Bilayer Hexagonal Boron Nitride}}},\ }\href {https://doi.org/10.1103/PhysRevLett.111.036104} {\bibfield  {journal} {\bibinfo  {journal} {Phys. Rev. Lett.}\ }\textbf {\bibinfo {volume} {111}},\ \bibinfo {pages} {036104} (\bibinfo {year} {2013})}\BibitemShut {NoStop}%
\bibitem [{\citenamefont {Pease}(1950)}]{pease_crystal_1950}%
  \BibitemOpen
  \bibfield  {author} {\bibinfo {author} {\bibfnamefont {R.~S.}\ \bibnamefont {Pease}},\ }\bibfield  {title} {\bibinfo {title} {Crystal {{Structure}} of {{Boron Nitride}}},\ }\href {https://doi.org/10.1038/165722b0} {\bibfield  {journal} {\bibinfo  {journal} {Nature}\ }\textbf {\bibinfo {volume} {165}},\ \bibinfo {pages} {722} (\bibinfo {year} {1950})}\BibitemShut {NoStop}%
\bibitem [{\citenamefont {Cavalcante}\ \emph {et~al.}(2018)\citenamefont {Cavalcante}, \citenamefont {{da Costa}}, \citenamefont {Farias}, \citenamefont {Reichman},\ and\ \citenamefont {Chaves}}]{cavalcante_stark_2018}%
  \BibitemOpen
  \bibfield  {author} {\bibinfo {author} {\bibfnamefont {L.~S.~R.}\ \bibnamefont {Cavalcante}}, \bibinfo {author} {\bibfnamefont {D.~R.}\ \bibnamefont {{da Costa}}}, \bibinfo {author} {\bibfnamefont {G.~A.}\ \bibnamefont {Farias}}, \bibinfo {author} {\bibfnamefont {D.~R.}\ \bibnamefont {Reichman}},\ and\ \bibinfo {author} {\bibfnamefont {A.}~\bibnamefont {Chaves}},\ }\bibfield  {title} {\bibinfo {title} {Stark shift of excitons and trions in two-dimensional materials},\ }\href {https://doi.org/10.1103/PhysRevB.98.245309} {\bibfield  {journal} {\bibinfo  {journal} {Phys. Rev. B}\ }\textbf {\bibinfo {volume} {98}},\ \bibinfo {pages} {245309} (\bibinfo {year} {2018})}\BibitemShut {NoStop}%
\bibitem [{\citenamefont {Kiper}\ \emph {et~al.}(2025)\citenamefont {Kiper}, \citenamefont {Adlong}, \citenamefont {Christianen}, \citenamefont {Kroner}, \citenamefont {Watanabe}, \citenamefont {Taniguchi},\ and\ \citenamefont {{\.I}mamo{\u g}lu}}]{kiper_confined_2025}%
  \BibitemOpen
  \bibfield  {author} {\bibinfo {author} {\bibfnamefont {N.}~\bibnamefont {Kiper}}, \bibinfo {author} {\bibfnamefont {H.~S.}\ \bibnamefont {Adlong}}, \bibinfo {author} {\bibfnamefont {A.}~\bibnamefont {Christianen}}, \bibinfo {author} {\bibfnamefont {M.}~\bibnamefont {Kroner}}, \bibinfo {author} {\bibfnamefont {K.}~\bibnamefont {Watanabe}}, \bibinfo {author} {\bibfnamefont {T.}~\bibnamefont {Taniguchi}},\ and\ \bibinfo {author} {\bibfnamefont {A.}~\bibnamefont {{\.I}mamo{\u g}lu}},\ }\bibfield  {title} {\bibinfo {title} {Confined {{Trions}} and {{Mott-Wigner States}} in a {{Purely Electrostatic Moir}}{\textbackslash}'e {{Potential}}},\ }\href {https://doi.org/10.1103/PhysRevX.15.011049} {\bibfield  {journal} {\bibinfo  {journal} {Phys. Rev. X}\ }\textbf {\bibinfo {volume} {15}},\ \bibinfo {pages} {011049} (\bibinfo {year} {2025})}\BibitemShut {NoStop}%
\bibitem [{\citenamefont {Stier}\ \emph {et~al.}(2018)\citenamefont {Stier}, \citenamefont {Wilson}, \citenamefont {Velizhanin}, \citenamefont {Kono}, \citenamefont {Xu},\ and\ \citenamefont {Crooker}}]{stier_magnetooptics_2018}%
  \BibitemOpen
  \bibfield  {author} {\bibinfo {author} {\bibfnamefont {A.~V.}\ \bibnamefont {Stier}}, \bibinfo {author} {\bibfnamefont {N.~P.}\ \bibnamefont {Wilson}}, \bibinfo {author} {\bibfnamefont {K.~A.}\ \bibnamefont {Velizhanin}}, \bibinfo {author} {\bibfnamefont {J.}~\bibnamefont {Kono}}, \bibinfo {author} {\bibfnamefont {X.}~\bibnamefont {Xu}},\ and\ \bibinfo {author} {\bibfnamefont {S.~A.}\ \bibnamefont {Crooker}},\ }\bibfield  {title} {\bibinfo {title} {Magnetooptics of {{Exciton Rydberg States}} in a {{Monolayer Semiconductor}}},\ }\href {https://doi.org/10.1103/PhysRevLett.120.057405} {\bibfield  {journal} {\bibinfo  {journal} {Phys. Rev. Lett.}\ }\textbf {\bibinfo {volume} {120}},\ \bibinfo {pages} {057405} (\bibinfo {year} {2018})}\BibitemShut {NoStop}%
\bibitem [{\citenamefont {Ben~Mhenni}\ \emph {et~al.}(2025)\citenamefont {Ben~Mhenni}, \citenamefont {Van~Tuan}, \citenamefont {Geilen}, \citenamefont {Petri{\'c}}, \citenamefont {Erdi}, \citenamefont {Watanabe}, \citenamefont {Taniguchi}, \citenamefont {Tongay}, \citenamefont {M{\"u}ller}, \citenamefont {Wilson}, \citenamefont {Finley}, \citenamefont {Dery},\ and\ \citenamefont {Barbone}}]{ben_mhenni_breakdown_2025}%
  \BibitemOpen
  \bibfield  {author} {\bibinfo {author} {\bibfnamefont {A.}~\bibnamefont {Ben~Mhenni}}, \bibinfo {author} {\bibfnamefont {D.}~\bibnamefont {Van~Tuan}}, \bibinfo {author} {\bibfnamefont {L.}~\bibnamefont {Geilen}}, \bibinfo {author} {\bibfnamefont {M.~M.}\ \bibnamefont {Petri{\'c}}}, \bibinfo {author} {\bibfnamefont {M.}~\bibnamefont {Erdi}}, \bibinfo {author} {\bibfnamefont {K.}~\bibnamefont {Watanabe}}, \bibinfo {author} {\bibfnamefont {T.}~\bibnamefont {Taniguchi}}, \bibinfo {author} {\bibfnamefont {S.~A.}\ \bibnamefont {Tongay}}, \bibinfo {author} {\bibfnamefont {K.}~\bibnamefont {M{\"u}ller}}, \bibinfo {author} {\bibfnamefont {N.~P.}\ \bibnamefont {Wilson}}, \bibinfo {author} {\bibfnamefont {J.~J.}\ \bibnamefont {Finley}}, \bibinfo {author} {\bibfnamefont {H.}~\bibnamefont {Dery}},\ and\ \bibinfo {author} {\bibfnamefont {M.}~\bibnamefont {Barbone}},\ }\bibfield  {title} {\bibinfo {title} {Breakdown of the {{Static Dielectric Screening Approximation}} of {{Coulomb Interactions}} in {{Atomically Thin
  Semiconductors}}},\ }\href {https://doi.org/10.1021/acsnano.4c11563} {\bibfield  {journal} {\bibinfo  {journal} {ACS Nano}\ }\textbf {\bibinfo {volume} {19}},\ \bibinfo {pages} {4269} (\bibinfo {year} {2025})}\BibitemShut {NoStop}%
\bibitem [{\citenamefont {Kim}\ \emph {et~al.}(2025)\citenamefont {Kim}, \citenamefont {Dery},\ and\ \citenamefont {Van~Tuan}}]{kim_excitons_2025}%
  \BibitemOpen
  \bibfield  {author} {\bibinfo {author} {\bibfnamefont {J.}~\bibnamefont {Kim}}, \bibinfo {author} {\bibfnamefont {H.}~\bibnamefont {Dery}},\ and\ \bibinfo {author} {\bibfnamefont {D.}~\bibnamefont {Van~Tuan}},\ }\bibfield  {title} {\bibinfo {title} {Excitons in fractionally filled moir{\textbackslash}'e superlattices},\ }\href {https://doi.org/10.1103/xdcj-cksl} {\bibfield  {journal} {\bibinfo  {journal} {Phys. Rev. B}\ }\textbf {\bibinfo {volume} {112}},\ \bibinfo {pages} {L041301} (\bibinfo {year} {2025})}\BibitemShut {NoStop}%
\bibitem [{\citenamefont {Liu}\ \emph {et~al.}(2021)\citenamefont {Liu}, \citenamefont {{van Baren}}, \citenamefont {Lu}, \citenamefont {Taniguchi}, \citenamefont {Watanabe}, \citenamefont {Smirnov}, \citenamefont {Chang},\ and\ \citenamefont {Lui}}]{liu_exciton-polaron_2021}%
  \BibitemOpen
  \bibfield  {author} {\bibinfo {author} {\bibfnamefont {E.}~\bibnamefont {Liu}}, \bibinfo {author} {\bibfnamefont {J.}~\bibnamefont {{van Baren}}}, \bibinfo {author} {\bibfnamefont {Z.}~\bibnamefont {Lu}}, \bibinfo {author} {\bibfnamefont {T.}~\bibnamefont {Taniguchi}}, \bibinfo {author} {\bibfnamefont {K.}~\bibnamefont {Watanabe}}, \bibinfo {author} {\bibfnamefont {D.}~\bibnamefont {Smirnov}}, \bibinfo {author} {\bibfnamefont {Y.-C.}\ \bibnamefont {Chang}},\ and\ \bibinfo {author} {\bibfnamefont {C.~H.}\ \bibnamefont {Lui}},\ }\bibfield  {title} {\bibinfo {title} {Exciton-polaron {{Rydberg}} states in monolayer {{MoSe2}} and {{WSe2}}},\ }\href {https://doi.org/10.1038/s41467-021-26304-w} {\bibfield  {journal} {\bibinfo  {journal} {Nat Commun}\ }\textbf {\bibinfo {volume} {12}},\ \bibinfo {pages} {6131} (\bibinfo {year} {2021})}\BibitemShut {NoStop}%
\bibitem [{\citenamefont {T{\"u}{\u g}en}\ \emph {et~al.}(2025)\citenamefont {T{\"u}{\u g}en}, \citenamefont {Seiler}, \citenamefont {Watanabe}, \citenamefont {Taniguchi}, \citenamefont {Kroner},\ and\ \citenamefont {{\.I}mamo{\u g}lu}}]{tugen_optical_2025}%
  \BibitemOpen
  \bibfield  {author} {\bibinfo {author} {\bibfnamefont {A.}~\bibnamefont {T{\"u}{\u g}en}}, \bibinfo {author} {\bibfnamefont {A.~M.}\ \bibnamefont {Seiler}}, \bibinfo {author} {\bibfnamefont {K.}~\bibnamefont {Watanabe}}, \bibinfo {author} {\bibfnamefont {T.}~\bibnamefont {Taniguchi}}, \bibinfo {author} {\bibfnamefont {M.}~\bibnamefont {Kroner}},\ and\ \bibinfo {author} {\bibfnamefont {A.}~\bibnamefont {{\.I}mamo{\u g}lu}},\ }\href {https://doi.org/10.48550/arXiv.2506.06098} {\bibinfo {title} {Optical {{Injection}} and {{Detection}} of {{Long-Lived Interlayer Excitons}} in van der {{Waals Heterostructures}}}} (\bibinfo {year} {2025}),\ \Eprint {https://arxiv.org/abs/2506.06098} {arXiv:2506.06098 [cond-mat]} \BibitemShut {NoStop}%
\bibitem [{\citenamefont {Carbone}\ \emph {et~al.}(2025)\citenamefont {Carbone}, \citenamefont {Breev}, \citenamefont {Figueiredo}, \citenamefont {Kretschmer}, \citenamefont {Geilen}, \citenamefont {Ben~Mhenni}, \citenamefont {Arceri}, \citenamefont {Krasheninnikov}, \citenamefont {Wubs}, \citenamefont {Holleitner}, \citenamefont {Huck}, \citenamefont {Kastl},\ and\ \citenamefont {Stenger}}]{carbone_quantifying_2025}%
  \BibitemOpen
  \bibfield  {author} {\bibinfo {author} {\bibfnamefont {A.}~\bibnamefont {Carbone}}, \bibinfo {author} {\bibfnamefont {I.~D.}\ \bibnamefont {Breev}}, \bibinfo {author} {\bibfnamefont {J.}~\bibnamefont {Figueiredo}}, \bibinfo {author} {\bibfnamefont {S.}~\bibnamefont {Kretschmer}}, \bibinfo {author} {\bibfnamefont {L.}~\bibnamefont {Geilen}}, \bibinfo {author} {\bibfnamefont {A.}~\bibnamefont {Ben~Mhenni}}, \bibinfo {author} {\bibfnamefont {J.}~\bibnamefont {Arceri}}, \bibinfo {author} {\bibfnamefont {A.~V.}\ \bibnamefont {Krasheninnikov}}, \bibinfo {author} {\bibfnamefont {M.}~\bibnamefont {Wubs}}, \bibinfo {author} {\bibfnamefont {A.~W.}\ \bibnamefont {Holleitner}}, \bibinfo {author} {\bibfnamefont {A.}~\bibnamefont {Huck}}, \bibinfo {author} {\bibfnamefont {C.}~\bibnamefont {Kastl}},\ and\ \bibinfo {author} {\bibfnamefont {N.}~\bibnamefont {Stenger}},\ }\bibfield  {title} {\bibinfo {title} {Quantifying the creation of negatively charged boron vacancies in {{He-ion}} irradiated hexagonal boron nitride},\
  }\href {https://doi.org/10.1103/PhysRevMaterials.9.056203} {\bibfield  {journal} {\bibinfo  {journal} {Phys. Rev. Materials}\ }\textbf {\bibinfo {volume} {9}},\ \bibinfo {pages} {056203} (\bibinfo {year} {2025})}\BibitemShut {NoStop}%
\end{thebibliography}%

\section*{\label{sec:acknowledgements}Acknowledgments}
A.B.M. acknowledges funding from the International Max Planck Research School for Quantum Science and Technology (IMPRS-QST) and from the Deutsche Forschungsgemeinschaft (DFG, German Research Foundation) via Germany’s Excellence Strategy (MCQST, EXC-2111/390814868).
K.W. and T.T. acknowledge support from the JSPS KAKENHI (Grant Numbers 21H05233 and 23H02052) , the CREST (JPMJCR24A5), JST and World Premier International Research Center Initiative (WPI), MEXT, Japan.

\section*{\label{sec:contributions}Author contributions}
A.B.M. conceived the project. A.B.M. and N.P.W. managed the research. A.B.M. and E.C. fabricated the devices. A.B.M. and E.C. performed the optical measurements. A.B.M., N.P.W., and J.J.F. analyzed the results. K.W. and T.T. grew bulk hBN crystals. A.B.M. and N.P.W. wrote the manuscript with input from all the authors.

\section*{\label{sec:interests}Competing interests}
The authors declare no competing interests.

\onecolumngrid
\renewcommand{\figurename}{\textbf{Extended Data Figure}}  
\setcounter{figure}{0}    

\newpage
\section*{\label{sec:si}Extended data}

\begin{figure*}[!ht]
\includegraphics[width=1\textwidth]{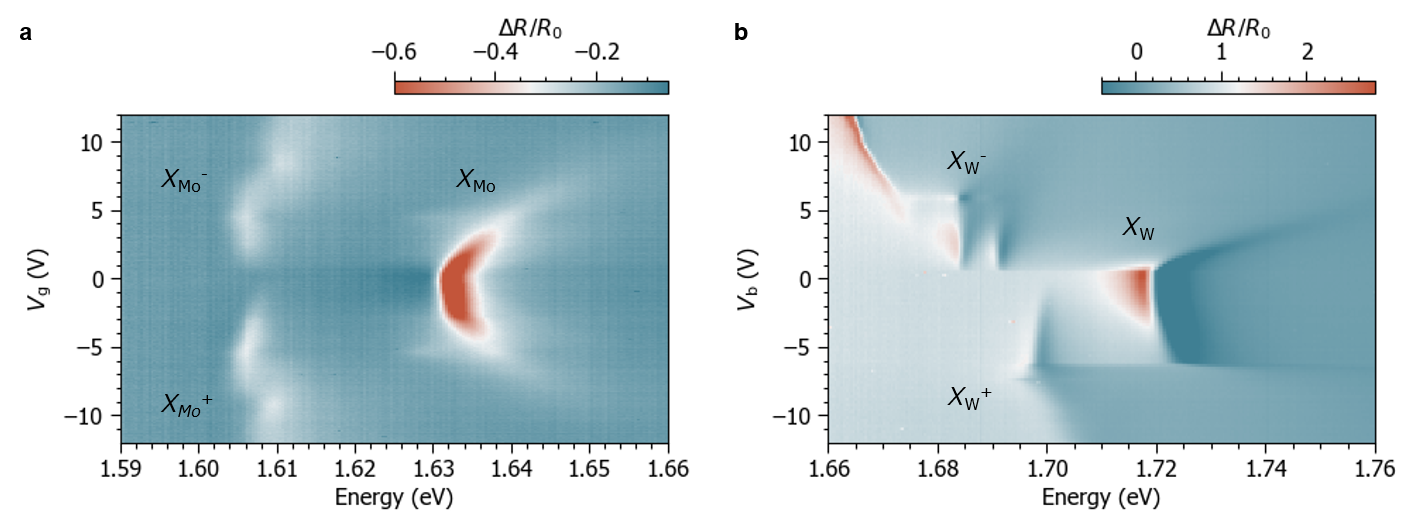}
\caption{\label{figext:monolayers}\textbf{TMD monolayers subject to a purely electrostatic \moire superlattice.}
\newline
MoSe2 (\textbf{a}) and WSe$_{2}$ monolayers (\textbf{b}) subject to a purely electrostatic \moire superlattice from a proximal twisted hBN bilayer.
The emergence of cusps is consistent with strongly correlated electron phases.
}
\end{figure*}
\newpage

\begin{figure*}[!ht]
\includegraphics[width=1\textwidth]{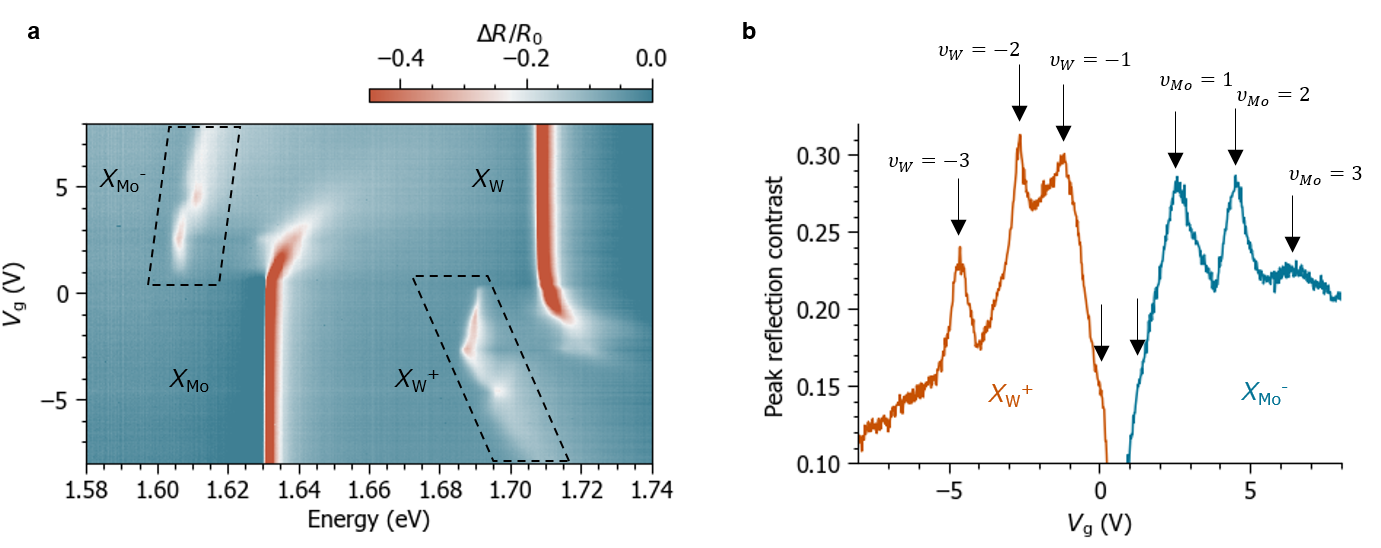}
\caption{\label{figext:analysis_device1}\textbf{Analysis of the strongly correlated electron phases in Device I.}
\newline
\textbf{a,} Shows the gate-dependent reflection contrast from Fig.~\ref{fig:device}e of the main text. The peak reflection contrast of $X_\text{Mo}^{-}$ (blue) and of $X_\text{W}^{+}$ (red) is extracted from the dashed regions in (a). 
The abrupt changes in the reflection contrast signal the emergence of the strongly correlated states at integer electron and hole fillings of the moiré superlattices.
Additionally, less pronounced dents signal the GWCs at $\nu _\text{Mo} = -1/3$ and $\nu _\text{W} = 1/3$.
}
\end{figure*}

\newpage

\begin{figure*}[!ht]
\includegraphics[width=1\textwidth]{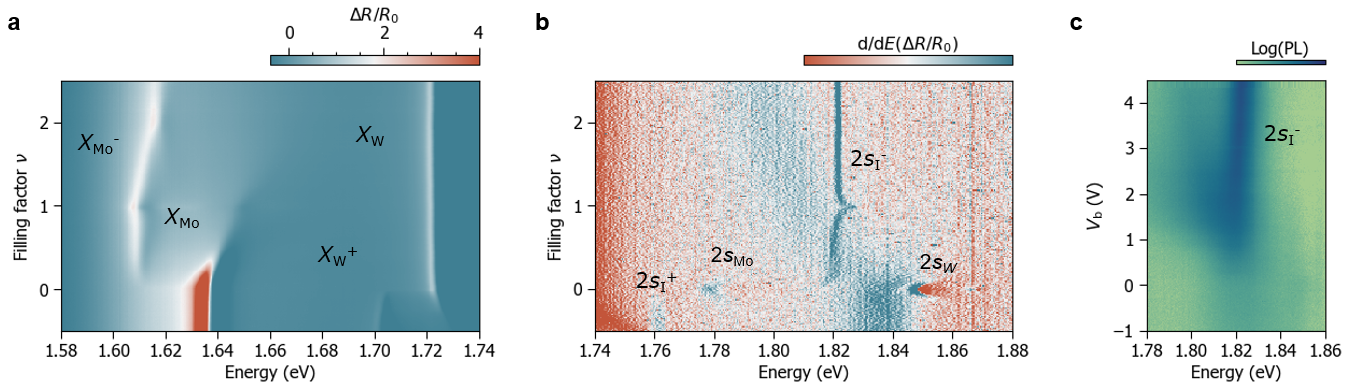}
\caption{\label{figext:rydberg}\textbf{Rydberg interlayer trions.}
\newline
\textbf{a,} Gate-dependent reflection contrast from Fig.~\ref{fig:mott}b, here including the WSe$_{2}$ excitons.
\textbf{b,} First derivative with respect to energy of the gate-dependent reflection contrast sweep from Fig.~\ref{fig:mott}c, here including the MoSe$_{2}$ Rydberg excitons.
In particular, the Rydberg interlayer trion $2s_\text{I}^\text{+}$ is also visible.
\textbf{c,} Rydberg interlayer trion from the control device with the natural (non-twisted) bilayer hBN.
}
\end{figure*}

\newpage

\begin{figure*}[!ht]
\includegraphics[width=1\textwidth]{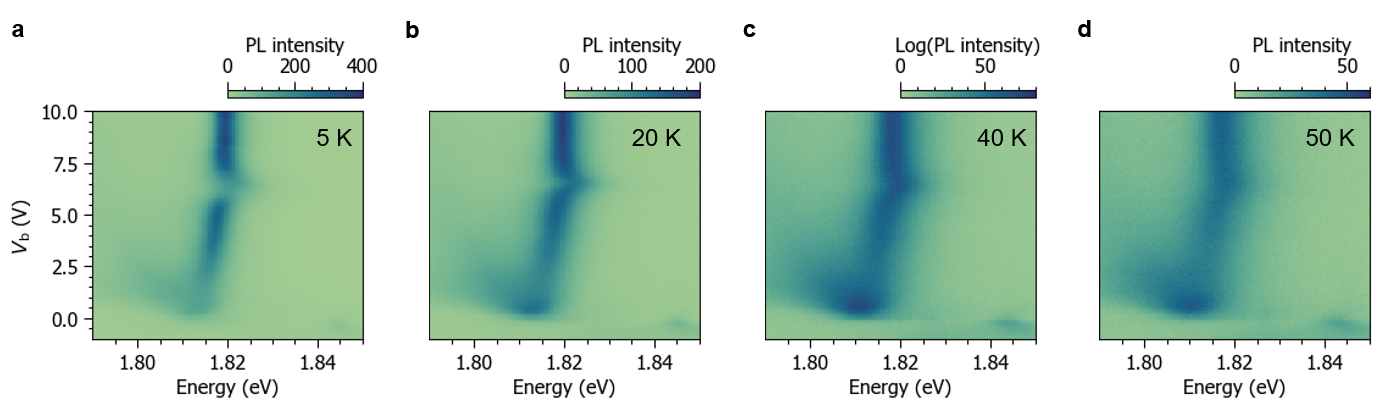}
\caption{\label{figext:melting_device2}\textbf{Melting of the electron Mott insulating phase from Device II.}
\newline
\textbf{a-d,} Gate-dependent PL sweeps showing the $2s_\text{I}^\text{-}$ from Device II at $5$ K (a), $20$ K (b), $40$ K (c), $50$ K (d).
}
\end{figure*}

\newpage

\begin{figure*}[!ht]
\includegraphics[width=1\textwidth]{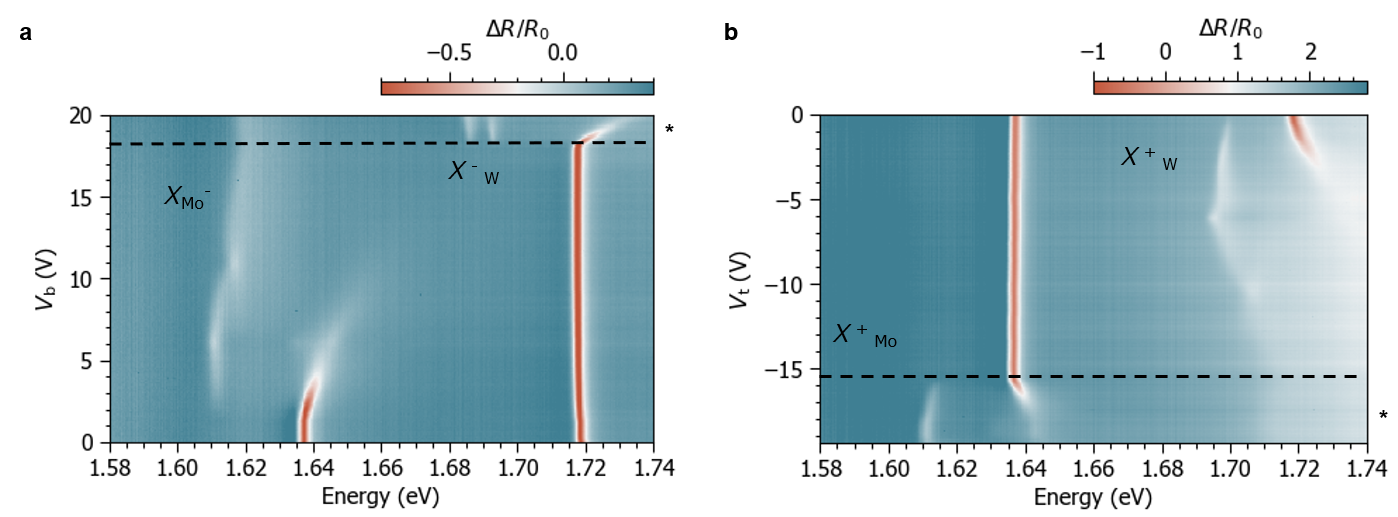}
\caption{\label{figext:same_type}\textbf{Simultaneous injection of the same carrier type in both layers of the dual \moire system.}
\newline
Gate-dependent reflection contrast sweep when electrons (\textbf{a}) or holes (\textbf{b}) are simultaneously injected into both MoSe$_{2}$ and WSe$_{2}$ layers.
This is evident from the coexistence of $X_\text{Mo}^{-}$ and $X_\text{W}^{-}$ in (a), and $X_\text{Mo}^{+}$ and $X_\text{W}^{+}$ in (b).
}
\end{figure*}

\newpage

\begin{figure*}[!ht]
\includegraphics[width=1\textwidth]{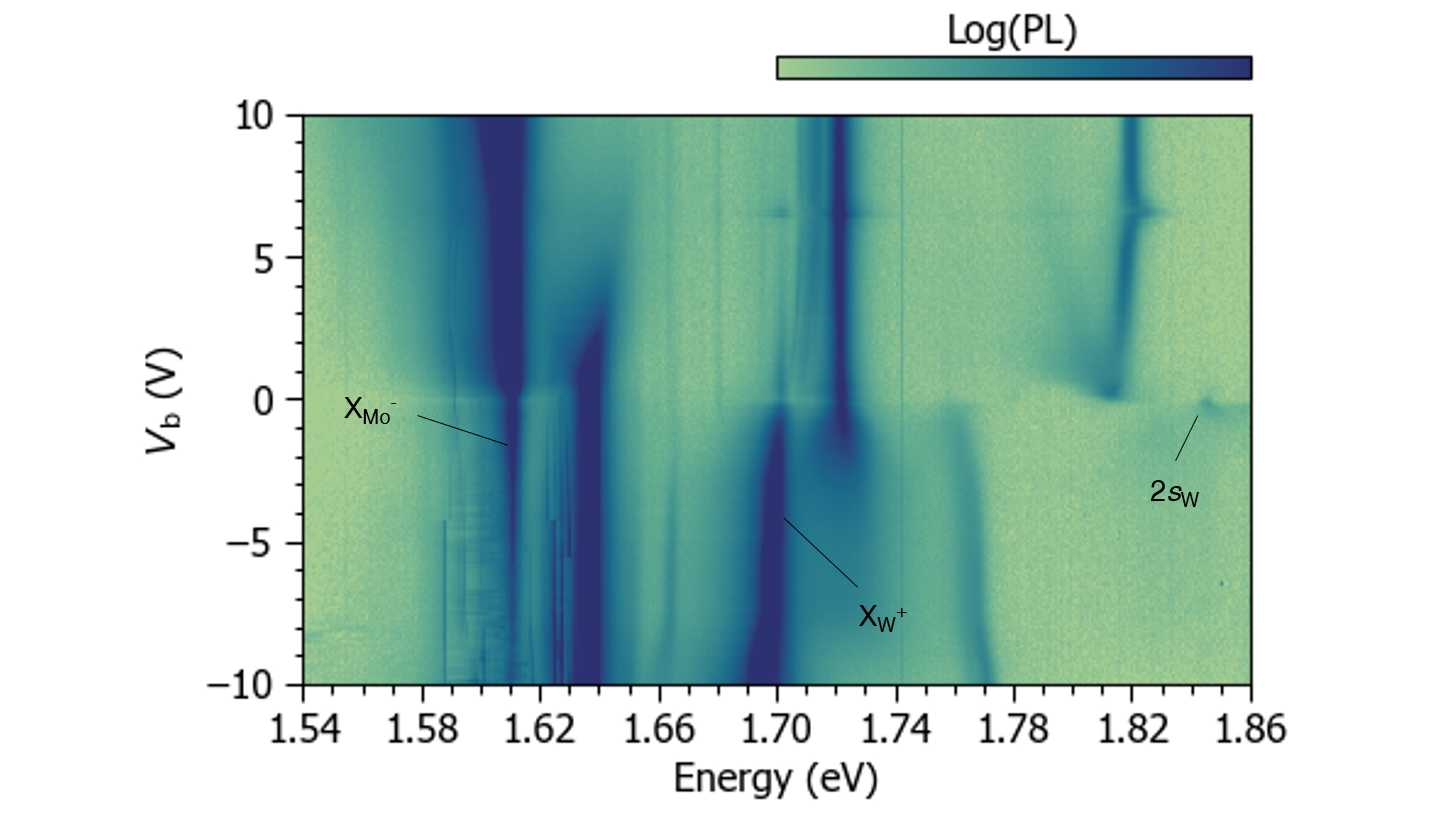}
\caption{\label{figext:dipolar}\textbf{Signature of dipolar excitons from Device II.}
\newline
Gate-dependent PL sweep showing the coexistence of $X_\text{Mo}^{-}$ in MoSe$_{2}$, the $X_\text{W}^{+}$ in WSe$_{2}$, and $2s_\text{W}$ in Device II.
}
\end{figure*}

\end{document}